\documentclass[A4paper,11pt]{article}
\pdfoutput=1
\usepackage{amsmath,amssymb}
\usepackage{graphicx,color}

\usepackage{jheppub} 

  \makeatletter
  \@addtoreset{equation}{section}
  \makeatother

\newcommand{\beq}{\begin{equation}}
\newcommand{\eeq}{\end{equation}}
\newcommand{\beqa}{\begin{eqnarray}}
\newcommand{\eeqa}{\end{eqnarray}}
\newcommand{\bea}{\begin{eqnarray}}
\newcommand{\eea}{\end{eqnarray}}

\newcommand{\nn}{\nonumber}

\newcommand{\ie}{{\it i.e.,\,}}
\newcommand{\eg}{{\it e.g.,\,}}

\newcommand{\lp}{\left(}
\newcommand{\rp}{\right)}

\newcommand{\ord}[1]{{\mathcal O}\lp #1\rp}

\def\v2#1{{\color{blue}{#1}}}

\newcommand{\fr}[1]{\frac{1}{#1}}

\newcommand{\Ord}[1]{{\mathcal O}\left(#1\right)}

\newcommand{\cD}{{\mathcal D}}

\newcommand{\nonum}{\nonumber\\ }

\newcommand{\vrho}{{\varrho}}

\newcommand{\cout}[1]{}

\begin{document}

\title{\boldmath Topology-changing horizons at large $D$ as Ricci flows}


\author[a,b]{Roberto Emparan}
\author[b,c]{and Ryotaku Suzuki}


\affiliation[a]{Instituci\'o Catalana de Recerca i Estudis Avan\c cats (ICREA),\\
Passeig Llu\'is Companys 23, E-08010, Barcelona, Spain}
\affiliation[b]{Departament de F\'isica Qu\`antica i Astrof\'isica, Institut de Ci\`encies del Cosmos, \\Universitat de Barcelona, Mart\'i i Franqu\`es 1, E-08028 Barcelona, Spain}
\affiliation[c]{Department of Physics, Osaka City University,\\
Sugimoto 3-3-138, Osaka 558-8585, Japan}

\emailAdd{emparan@ub.edu}
\emailAdd{s.ryotaku@icc.ub.edu}

\abstract{
The topology-changing transition between black strings and black holes localized in a Kaluza-Klein circle is investigated in an expansion in the inverse of the number of dimensions $D$. Performing a new kind of large-$D$ scaling reduces the problem to a Ricci flow of the near-horizon geometry as it varies along the circle direction. The flows of interest here simplify to a non-linear logarithmic diffusion equation, with solutions known in the literature which are interpreted as the smoothed conifold geometries involved in the transition, namely, \textit{split} and \textit{fused} cones, which connect to black holes and non-uniform black strings away from the conical region. Our study demonstrates the adaptability of the $1/D$ expansion to deal with all the regimes and aspects of the static black hole/black string system, and provides another instance of the manner in which the large $D$ limit reduces the task of solving Einstein's equations to a simpler but compelling mathematical problem.
}

\maketitle

\section{Introduction}

The study of black strings and black holes in a spacetime with a compactified  circle has been a fertile source of insights into the physics of black hole horizons\footnote{See \cite{Kol:2004ww,Harmark:2007md} for early reviews, and the book monograph \cite{Horowitz:2012nnc} for a more up to date view.}. The discovery by Gregory and Laflamme \cite{Gregory:1993vy} that black strings in a long enough circle are unstable to small perturbations prompted two lines of investigation, both of them eventually involving, although in different manners, the appearance of a naked singularity on the horizon and a change in horizon topology when this singularity is resolved. The first line, which is the one that we pursue in this article, studies the static solutions of black holes and strings in a Kaluza-Klein circle and the connections between them. The other line aims at following the non-linear dynamical evolution of the instability, a distinct problem that requires very different methods \cite{Lehner:2010pn,Horowitz:2012nnc}.

The small static deformation of black strings at the threshold of the Gregory-Laflamme instability leads to a branch of non-uniform black string solutions. Following them as their inhomogeneity grows, one arrives at a configuration in which the black string horizon pinches to zero thickness at a singular point along its length. This configuration mediates a change in the topology of the horizon: the black strings with horizon topology $S^1\times S^{D-3}$ give way to black holes localized in the circle, with $S^{D-2}$ topology.\footnote{The evolution of black strings to black holes in solution space seems to be unrelated to the dynamical evolution of black strings towards (and across) a naked singularity. Although the change in topology is expected to be the same in both cases, the dynamical evolution appears to be at all moments strongly time-dependent, and occurs far from the static configurations around the transition in solution space. The singularities also appear to be very different in the two cases.} Moving in the opposite direction, we have a merger transition from disconnected black hole horizons (the periodic images of a black hole in a circle) to the continuously connected horizon of a black string.\footnote{Again, these black hole mergers in solution space are rather different than the dynamical mergers of event horizons. The latter admit simple local models studied in \cite{Emparan:2017vyp} and do not involve naked singularities. The evolution in solution space is adiabatic, while the dynamical merger is irreversible.}

\begin{figure}[t]
\begin{center}
\includegraphics[width=.8\linewidth]{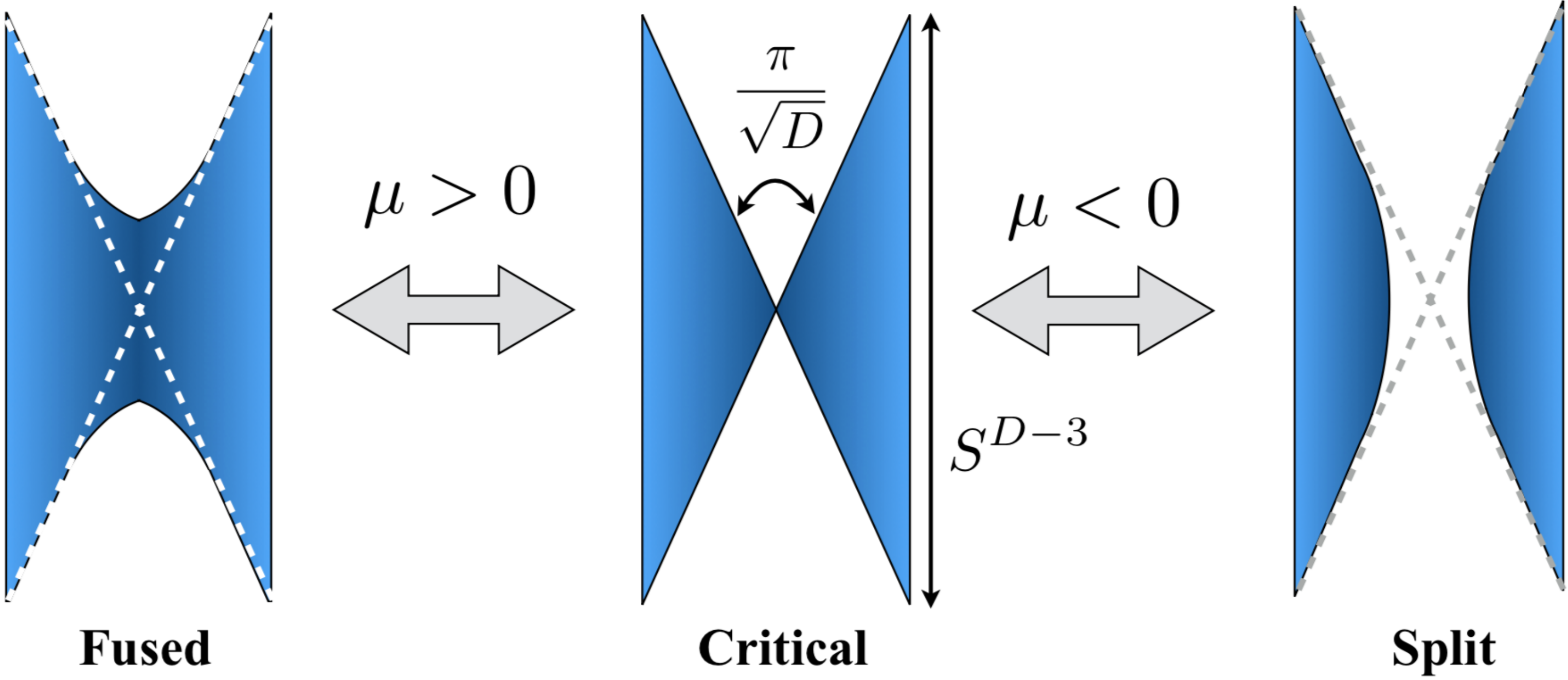}
\caption{Topology change across the black string/black hole transition, as first proposed in \cite{Kol:2002xz}.  $\mu$ is the parameter in the solutions that changes sign across the transition.}
\label{fig:LDcone_topologychange}
\end{center}
\end{figure}

The singular configurations at the topological transition are interesting. In ref.~\cite{Kol:2002xz}, Kol proposed an exact local model for the region near the singularity. Its Euclideanized geometry is a self-similar cone over $S^2\times S^{D-3}$, and the black hole/black string transition is akin to the conifold transitions that appear in Calabi-Yau spaces, which connect two topologically different phases. In the present case, on one side of the transition we have geometries in which the $S^2$ does never shrink to zero. These correspond, upon Wick-rotating an angle of the $S^2$, to Lorentzian `black hole phases' with two disconnected horizons, and therefore we will refer to them as \textit{split cones}.
At the other side of the transition, the $S^{D-3}$ remains of finite size throughout the horizon, and we obtain the region near the neck of `black string phases'; we call these \textit{fused cones} (see figure \ref{fig:LDcone_topologychange}). Note that the terms `split' and `fused' are chosen for the shape of the cone over $S^{D-3}$, \ie for the horizon in the Lorentzian geometries. Related critical cone geometries were argued to universally control the topology change in other systems of higher-dimensional black holes, \textit{e.g.}, black hole-black ring transitions \cite{Emparan:2011ve}.

\begin{figure}[t]
\begin{center}
\includegraphics[width=.8\linewidth]{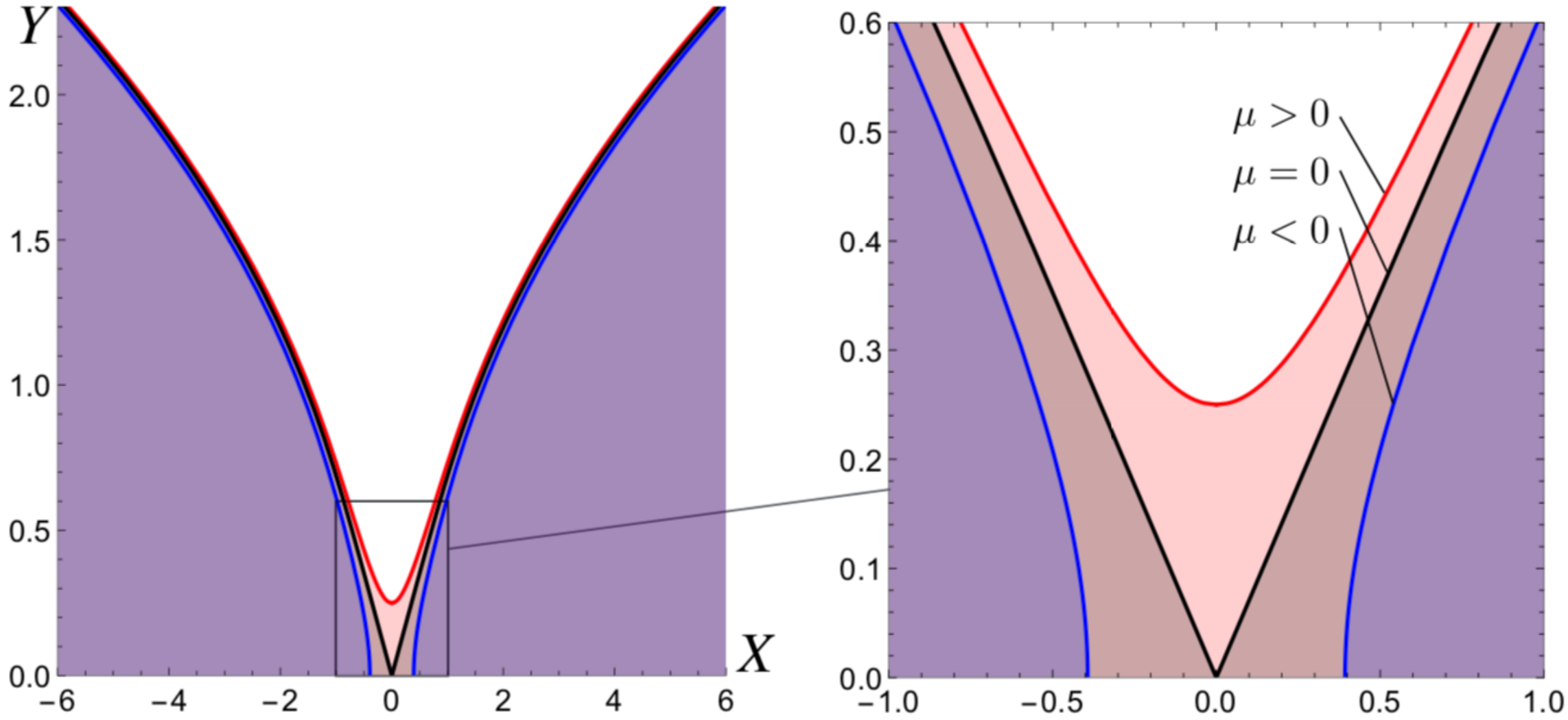}
\caption{Geometry of the horizon near the merger transition as obtained in this article (cf.~eqs.~\eqref{eq:horizonds2} and \eqref{eq:flatembed}). The black curve
is  the critical solution ($\mu=0$), the blue curve the localized black hole ($\mu=-1/16$), the red curve the nonuniform black string ($\mu=1/16$). At short distances the singular cones are smoothed (split or fused) for $\mu\neq 0$. At large distances all these geometries deviate from the conical shape to connect to localized black hole and highly non-uniform black string horizons; both of these can be regarded, away from the conical region, as slightly deformed Schwarzschild black holes. The proper physical size of the magnified conical region (right) is $\ord{1/D}$ along the horizontal axis and $\ord{1/\sqrt{D}}$ along the vertical axis.}
\label{fig:embedding}
\end{center}
\end{figure}

Although the exactly self-similar conical geometry is simple, its split and fused deformations to the black hole and the black string phases are only known through approximate perturbative or numerical solutions \cite{Kol:2003ja,Kudoh:2004hs,Asnin:2006ip,Emparan:2014pra,Kalisch:2016fkm,Kalisch:2017bin,Cardona:2018shd,Ammon:2018sin}. Furthermore, the connection of the local model to the overall geometry of the black hole or black string horizon away from the self-similar region is poorly understood. In this paper we make progress in these two directions by giving a complete solution of this system in an expansion in $1/D$ \cite{Asnin:2007rw,Emparan:2013moa,Emparan:2013xia,Emparan:2014aba,Emparan:2015hwa,Emparan:2015gva,Bhattacharyya:2015dva,Bhattacharyya:2015fdk}. More concretely, this method allows us to study how the self-similarity of the conifold is broken in two regimes: at short distance, where the tip of the cone is deformed (split or fused), and at large distance, where the horizon asymptotically bends away from the conical shape and connects to the geometry of a slightly deformed Schwarzschild black hole. Our main results for the shape of the horizon are illustrated in figure \ref{fig:embedding}.

Our analysis also reveals a fascinating property of  horizon mergers in the large-$D$ limit: the near-horizon geometry of a very general class of them changes along the cone direction obeying the Ricci flow equation \cite{Hamilton1982}. 
In the black hole/black string system, Ricci flow makes the Euclidean $S^2$ factor of the geometry shrink. The flow governs the topology of the merger according to whether it collapses the $S^2$ (fused cone) or not (split cone). This two-dimensional flow actually reduces to a non-linear logarithmic diffusion equation, which has been extensively studied in the literature with known solutions that play an important role in the black hole-black string system. 

This study culminates the investigation of the black hole-black string problem in the large-$D$ expansion. Refs.~\cite{Emparan:2015hwa,Emparan:2015gva}, and especially \cite{Emparan:2018bmi}, developed this approach to obtain non-uniform black strings and their phase diagram in dimensions from $D\approx 9$ to $\infty$, by solving its equations up to fourth order in $1/D$. The method, however, breaks down before the singular pinch-off point is reached. The results in this paper, which not only comprise the singularity and its non-singular deformations, but also extend the solutions into the localized black hole phase, successfully complete the program to describe the black hole-black string system at large $D$.

Finally, it is interesting to note that our study involves a new kind of large-$D$ scaling limit. In particular, it is different than the one used in the effective theory of black holes and branes \cite{Emparan:2015hwa,Emparan:2015gva,Bhattacharyya:2015dva,Bhattacharyya:2015fdk}. We suspect that similar scalings may play a role in other critical systems.

\paragraph{Outline.}In the next section, we motivate the new type of large-$D$ scaling by studying the exact singular double cone geometries, and then solve the Einstein equations to leading order in $1/D$ to find critical, split and fused cones. In Section \ref{sec:ricciflow}, we derive the Ricci flow equation as the large-$D$ limit of the Einstein equations for general topology-changing transitions.  Section \ref{sec:2dmerger} contains the bulk of the analysis and results of the paper. In it, we show that the large-$D$ conifolds must solve the logarithmic diffusion equation, and study its solutions. The properties of the split conifold as a configuration of two gravitationally interacting black holes are analyzed in detail. Also, the geometry of the conifold away from the transition region is revealed to be a slightly deformed Schwarzschild solution, both for the fused and the split conifold.
In Section \ref{sec:neck-resolve}, we perform a finer resolution of the neck in fused cones which removes its singular behavior. Section \ref{sec:extensions} extends the Ricci flow equation to geometries with a cosmological constant or with scalar fields. Section \ref{sec:outlook} contains our concluding remarks.

\paragraph{Terminology}
\begin{itemize}
\item A \textit{cone} over a base $M$ is a geometry of the form
\begin{eqnarray}\label{eq:defcone}
 ds_{\rm cone}^2 = d\rho^2 + \rho^2 ds^2(M)\,,
\end{eqnarray}
at least asymptotically as $\rho\to\infty$.
Specifically, we will be considering \textit{double cones} with $M=S^p \times S^n$. Properly speaking, a cone has metric \eqref{eq:defcone} for all $\rho>0$, and is singular at the tip $\rho\to 0$. But we will also use the term cone for the geometries with asymptotically self-similar, conical shape as $\rho\to\infty$ but smoothed singularity at small $\rho$. These are \textit{split} or \textit{fused cones} (see figure \ref{fig:LDcone_topologychange}), corresponding to whether the Euclidean $S^p$ or the $S^n$ remain of finite size in the geometry. We refer to the self-similar singular geometry \eqref{eq:defcone} as a \textit{critical cone}.

\item A \textit{conifold} is a geometry that contains localized cone singularities, or localized regions that approach smoothed cones. In the conifolds in this paper, the $SO(p+1)$ symmetry of $S^p$ will be broken away from the conical region, but the $SO(n+1)$ of $S^n$ is preserved.

\end{itemize}

\section{Large-$D$ cones: critical, split and fused}\label{sec:cones}

\subsection{Exact critical cones}

The double cones in dimension
\beq
D=n+p+1,
\eeq
that were described in \cite{Kol:2002xz,Asnin:2006ip} and generalized in \cite{Emparan:2011ve}, are the metrics
\begin{eqnarray}
 ds^2 = d\rho^2 + \frac{p-1}{n+p-1}\rho^2 d\Omega_{p}^2+ \frac{n-1}{n+p-1} \rho^2 d\Omega^2_{n},\label{eq:doublecone}
\end{eqnarray}
where $d\Omega_p^2$ and $d\Omega_n^2$ are the line elements for $S^p$ and $S^n$ respectively. These are exact Ricci flat metrics for all $p,n\geq 2$, with a curvature singularity at the tip of the cone at $\rho=0$. The Lorentzian geometries obtained by Wick-rotating one azimuthal angle $\phi=it$ in $S^p$,
\beq
ds^2=d\rho^2 + \frac{p-1}{n+p-1}\rho^2 \lp-\cos^2\chi dt^2+d\chi^2+\sin^2\chi d\Omega_{p-2}\rp + \frac{n-1}{n+p-1} \rho^2 d\Omega^2_{n},\label{eq:lordoublecone}
\eeq
have horizons at the points where $\cos\chi=0$. When $p>2$ we have $\chi\in[0,\pi/2]$ and there is a single, connected horizon at $\chi=\pi/2$. Instead, when $p=2$ the polar angle $\chi$ runs in $[-\pi/2,\pi/2]$ and the interval endpoints correspond to two disconnected components of the horizon, which meet at $\rho=0$.

These geometries are not asymptotically flat, and the horizons extend indefinitely as $\rho\to\infty$. The conjecture that they arise locally at the connection between black string and black hole phases when $p=2$, and black ring and black hole phases when $p=3$, has been verified numerically in \cite{Kol:2003ja,Kudoh:2004hs,Emparan:2014pra,Kalisch:2016fkm,Kalisch:2017bin,Cardona:2018shd,Ammon:2018sin}. Ref.~\cite{Emparan:2014pra} also showed that $p=2$ mediates black hole-black Saturn transitions.

Here we wish to investigate this problem analytically. More precisely, we will study how the singularity in the conical solutions is resolved when either the $S^{p}$ or the $S^n$ are blown up at $\rho=0$, and how the solutions at large $\rho$ deviate from the conical shape to approach deformed Schwarzschild black holes. To this purpose, we resort to a study in the large $D$ limit, in which $n\to\infty$ while $p$ remains finite.

\subsection{A new kind of large-$D$ limit}

We expect that the resolution of the tip of the cone will preserve the $SO(p+1)\times SO(n+1)$ symmetry. Taking a cue from the exact double cone \eqref{eq:doublecone}, we consider the ansatz
\begin{eqnarray}
ds^2 = N^2(\rho) d\rho^2 + \fr{n}S(\rho)d\Omega_p^2+\rho^2 d\Omega_n^2,
\label{eq:largeDcone-ansatz}
\end{eqnarray}
with $n\simeq D\gg 1$ and $p=\ord{1}$. 
The horizon corresponds to the (Wick-rotated) poles of a rotational axis in $S^p$.  With our choice of radial gauge, the topology of the horizon depends on whether $S(\rho)$ and $\rho$ can reach zero values.

It is important to observe that the large-$D$ scaling in \eqref{eq:largeDcone-ansatz} is different than considered in the effective theory of black branes in \cite{Emparan:2015hwa,Emparan:2015gva,Bhattacharyya:2015dva,Bhattacharyya:2015fdk}. In that case, the coordinate orthogonal to the horizon is rescaled such that gradients in that direction are of size
\beq
\nabla_\perp=\Ord{D}\,.
\eeq 
This scaling is dictated by the properties of the Schwarzschild solution in the large-$D$ limit \cite{Emparan:2013xia}.
Instead, in the present case the double-cone geometry \eqref{eq:lordoublecone} instructs us to rescale the direction $\chi$ that points away from the horizon\footnote{The time direction is also rescaled, but trivially, as it is isometric.} so that orthogonal gradients are
\beq\label{eq:orthgrad}
\nabla_\perp=\Ord{\sqrt{D}}\,.
\eeq 
Therefore, the large-$D$ limit we are studying is a qualitatively new one. In section \ref{sec:single-BH} we will explain how this limit can deal with the near-horizon region of the black holes, which, in general, has radial extent $\sim 1/D$. Briefly, in this new limit this region is uniformly blown up by a factor $\sqrt{D}$ to size $\sim 1/\sqrt{D}$, so radial gradients are like \eqref{eq:orthgrad}.

\subsection{Smoothing the cone}\label{sec:smoothcones}

In the large $D$ limit, expanding in $1/n$, the Einstein equations become
\begin{eqnarray}
R_{ab} &\simeq& \fr{2}\left(2(p-1)-\frac{S'}{N^2\rho}\right)  \omega^{(p)}_{ab},\label{eq:largeDcone-eqlo_Sp}
\\ R_{ij} &\simeq& \frac{n}{\rho^2}\left(1-N^{-2}\right)\omega^{(n)}_{ij},\label{eq:largeDcone-eqlo_Sn}
\\ R_{\rho\rho} &\simeq& -\frac{n}{\rho} \partial_\rho \ln N,
\label{eq:largeDcone-eqlo_rho}
\end{eqnarray}
where $\omega^{(p)}_{ab}$ and $\omega^{(n)}_{ij}$ are the metrics for $S^p$  and $S^n$.
\cout{The leading order solution is
\begin{eqnarray}
B(\rho) = 1,\quad  S(\rho) = (p-1)(\rho^2-\mu),
\end{eqnarray}}
These equations are easily solved to find the leading order metric
\begin{eqnarray}
 ds^2 = d\rho^2 + \frac{p-1}{n}(\rho^2-\mu)d\Omega_p^2+\rho^2 d\Omega_n^2,
\label{eq:largeDcones}
\end{eqnarray}
where $\mu$ is an integration constant. These geometries have the same asymptotic conical structure at large $\rho$ as the double cones of \eqref{eq:doublecone}, but they allow for richer structure at small $\rho$.
Depending on the sign of the constant $\mu$, we obtain the following geometries (see figure \ref{fig:LDcone_topologychange}):
\begin{description}
\item[Fused cone: $\mu>0$]\mbox{} \\
$S(\rho)$ vanishes at a neck where $S^{n}$ reaches a minimum radius $\rho=\sqrt{\mu}$, which joins the two asymptotically conical regions into a single horizon of cylindrical topology $\mathbf{R}^{p-1}\times S^{n}$. The topology of the Euclidean manifold is $\mathbf{R}^{p+1}\times S^{n}$.
\item[Critical cone: $\mu=0$]\mbox{} \\
The double cone \eqref{eq:doublecone} is reproduced in the limit $n\to\infty$.
\item[Split cone: $\mu<0$]\mbox{} \\
$S(\rho)$ never vanishes but $\rho$ reaches down to zero. For $p=2$, the spacetime has two separate horizons, with minimum separation $\pi\sqrt{|\mu|/n}$. For $p>2$ the horizon topology is connected, $\mathbf{R}^{n+1}\times S^{p-2}$. The topology of the Euclidean manifold is $\mathbf{R}^{n+1}\times S^{p}$.
\end{description}
The split cone geometry is regular, but
the fused cone has a curvature singularity at the neck $\rho=\sqrt{\mu}$.
In section \ref{sec:neck-resolve}  we will show that this neck singularity is the result of poor resolution of this region in our expansion, and is resolved if we magnify it further and include $1/n$ corrections to the geometry.

One can show that the corrections to these solutions at any order in $1/n$ are normalizable in the region $\rho\to \infty$, so the asymptotic conical structure is not modified at any perturbative order. Therefore, if we want to study how the cones extend into a black hole or black string horizon at large $\rho$, we must resort to a more general class of solutions of which these cones are a limit.  We shall do this in the next sections.

\medskip

%
\section{Merging horizons as Ricci flows} \label{sec:ricciflow}

\begin{figure}[t]
\begin{center}
\includegraphics[width=.7\linewidth]{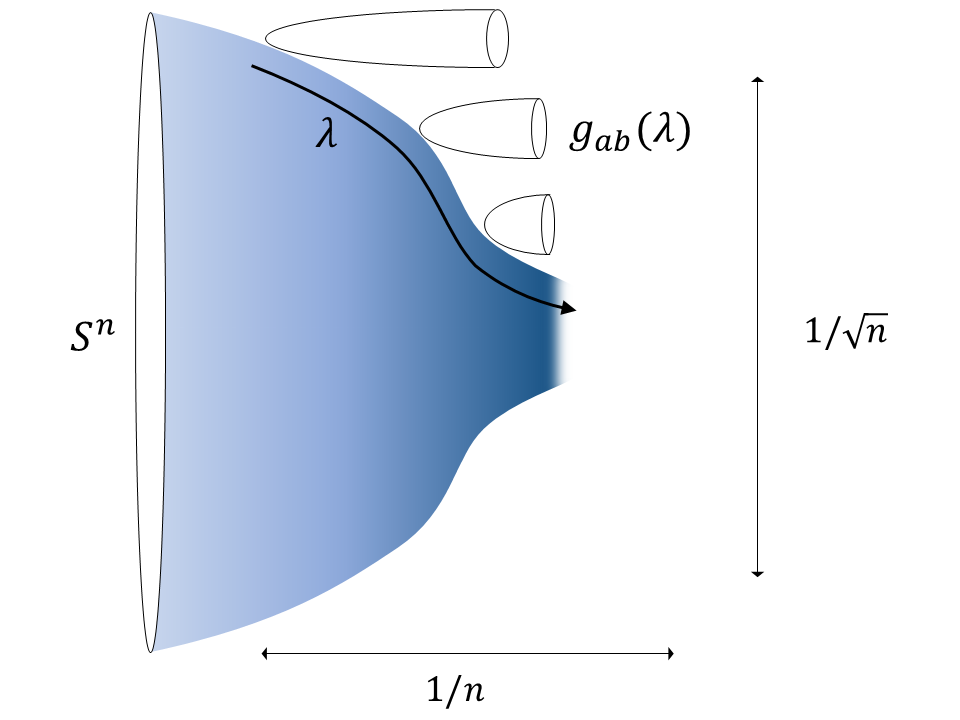}
\caption{Ricci flow of merger geometries in $D=p+n+1$ dimensions, in the limit $n\to\infty$. The $p$-dimensional metric $g_{ab}(\lambda)$ satisfies the Ricci flow equation with parameter $\lambda$, while the radius $\rho$ of the $S^n$ factor shrinks as $\rho=\sqrt{2(\lambda_0-\lambda)}$. In this article we discuss in detail two-dimensional flows in which the circular direction in (Euclidean) $g_{ab}$ shrinks to zero at the (Lorentzian) horizon of a black hole or black string. The flow starts at $\lambda=-\infty$, and depending on whether $g_{ab}$  collapses at $\lambda<\lambda_0$, at $\lambda=\lambda_0$, or does not collapse by $\lambda=\lambda_0$, we get fused, critical, or split conifolds, respectively.}
\label{fig:ricciflow}
\end{center}
\end{figure}

Now we consider product geometries with a $S^n$ factor with $SO(n+1)$ symmetry and $n\to\infty$, but we will not assume any symmetry (\eg $SO(p+1)$) in the $p$-dimensional factor. Specifically, we study geometries of the form\footnote{The more precise metric is \eqref{eq:riccflmetric} below.}
\begin{eqnarray}
ds^2 = N^2(\rho,y) d\rho^2 + \fr{n}g_{ab}(\rho,y) dy^a dy^b+\rho^2 d\Omega_n^2\,,
\end{eqnarray}
where $g_{ab}$ is a one-parameter family of $p$-dimensional metrics with parameter $\rho$, and we take $n\gg 1$. As in the cone metrics, we can regard $\rho$ as the coordinate orthogonal to $S^n$ along the horizon, or along the Kaluza-Klein circle in the complete configurations.
We will then show that if these geometries are to satisfy the Einstein vacuum equations to leading order at large-$D$, then they must solve the $p$-dimensional Ricci flow equation \cite{Hamilton1982},
\begin{eqnarray}
\frac{\partial g_{ab}}{\partial\lambda} = -2R_{ab}\,, \label{eq:ricci-flow-tau}
\end{eqnarray}
where the flow parameter is
\beq\label{eq:taurho}
\lambda=\lambda_0-\frac{\rho^2}{2}
\eeq
with $\lambda_0$ a constant, and $R_{ab}$ is the Ricci curvature of the metric $g_{ab}$ at constant $\rho$. We illustrate this in figure \ref{fig:ricciflow}. 

Therefore, the problem of finding the merger solution is recast as that of solving the Ricci flow from a given asymptotic geometry at $\lambda =-\infty$.\footnote{Ricci flow has been used in \cite{Headrick:2009pv} as a technique to solve numerically the Einstein equations for the black hole/black string system. However, we do not see any connection between those Ricci flows and ours.} In the Ricci flow literature, this is called the {\it ancient solution problem}, where the ancient solution is the Ricci flow solution defined back to $\lambda = -\infty$.

The topology of the flow depends on whether it collapses the metric $g_{ab}$ to zero before the $S^n$ shrinks to zero at $\lambda=\lambda_0$, or it collapses exactly at $\lambda=\lambda_0$, or it does not collapse by $\lambda=\lambda_0$. These possibilities correspond to fused, critical and split conifolds, respectively.

\subsection{Derivation}

We begin with a generalized version of the ansatz \eqref{eq:largeDcone-ansatz},\footnote{The $y$ dependence in $H$ can be eliminated. It would add a term $\bar{R} \sim n^3 (\nabla H)^2$ in the scalar constraint~\eqref{eq:adm-scalar-c}, which results in $\nabla H=0$.}
\begin{eqnarray}
ds^2 &=&N^2(\rho,x) d\rho^2 + \bar{g}_{AB}(\rho,x) dx^A dx^B \label{eq:ansatz-00}\\
&=&N^2(\rho,y) d\rho^2 + \fr{n}g_{ab}(\rho,y) dy^a dy^b+H^2(\rho)e^{2C(\rho,y)/n} d\Omega_n^2\,.
\label{eq:ansatz-0}
\end{eqnarray}
Bear in mind that, when matching this region to the entire geometry away from it, the metric must be multiplied by an overall factor of $1/n$. This accounts for the sizes in different directions shown in figure \ref{fig:ricciflow}: $\ord{1/n}$ in the directions $y$ orthogonal to $S^n$, and $\ord{1/\sqrt{n}}$ in $\rho$, away from the $S^n$ symmetry axis.

The vacuum Einstein equations for the metric \eqref{eq:ansatz-00} decompose into the evolution equation
\begin{eqnarray}
 \fr{N}\partial_\rho K^{A}{}_{B} = \bar{R}^{A}{}_{B}-KK^{A}{}_{B} - \fr{N}\bar{\nabla}^{A}\bar{\nabla}_{B} N,\label{eq:adm-evolve}
\end{eqnarray}
and the constraint equations
\begin{align}
&K^2-K^{A}{}_{B} K^{B}{}_{A} =\bar{R},\label{eq:adm-scalar-c} \\
&\bar{\nabla}_{B} K^{B}{}_{A}-\bar{\nabla}_{A}K = 0,\label{eq:adm-vector-c}
\end{align}
where the extrinsic curvature of surfaces of constant $\rho$ is
\begin{equation}
K_{AB} = \fr{2N} \partial_\rho \bar{g}_{AB}.
\end{equation}

From the scalar constraint~\eqref{eq:adm-scalar-c} we obtain
\begin{eqnarray}
 0=K^2-K^{A}{}_{B} K^{B}{}_{A} -\bar{R} = \frac{n^2}{H^2} \left[\frac{H'^2}{N^2} - 1\right]+\ord{n},
\end{eqnarray}
which we solve as
\beqa
 H &=&\rho,\\
 N &=& 1+\frac{N_1(\rho,y)}{n}.
\eeqa
This choice fixes the radial gauge to leading order in the large-$n$ expansion, but we can also conveniently fix it at the next order.
Up to $\ord{n^{-1}}$, we have the freedom to transform
\begin{equation}
 \rho \to \rho \left(1+\frac{\beta(\rho,y)}{n}\right),\quad y^a \to y^a + \xi^a(\rho,y),
\end{equation}
where $\xi^a$ is chosen to eliminate $g_{\rho a}$. This transformation changes the leading order of $g_{ab}$, $C$ and $N_1$, and we use it to set $C(\rho,y)=0$. As we will see below, this fixes $N_1$ on-shell.

With these choices the metric is
\begin{eqnarray}\label{eq:riccflmetric}
ds^2 = \left(1+\frac{2N_1(\rho,y)}{n}\right) d\rho^2 + \fr{n}g_{ab}(\rho,y) dy^a dy^b+\rho^2 d\Omega_n^2\,,
\end{eqnarray}
where we stress that the term with $N_1$, although apparently subleading in $1/n$, must be retained for consistency of the choice of radial gauge.

The extrinsic curvature up to $\ord{n^{-1}}$ is 
\begin{eqnarray}
 K_{ab} = \fr{2n}  \partial_\rho g_{ab},\quad K^i{}_j = \fr{\rho}\left(1-\frac{N_1}{n}\right)\delta^i{}_j,
\end{eqnarray}
where $i,j$ are directions along $S^n$,
and
\begin{eqnarray}
K = \frac{n}{\rho}\left[ 1+ \fr{n}\left(\fr{2}\rho\partial_\rho \ln g-N_1\right) \right].
\end{eqnarray}
The Ricci tensor is
\begin{eqnarray}
&& \bar{R}_{ab} = R_{ab}+\ord{n^{-1}},\\
 &&  \bar{R}^i{}_j =  \frac{n-1}{\rho^2} \delta^i{}_j +\ord{n^{-1}},
\end{eqnarray}
Combining these, the evolution equation~\eqref{eq:adm-evolve} for $K_{ab}$ becomes
\beq
K K_{ab}=R_{ab}+\ord{n^{-1}}
\eeq
\ie\ the second derivatives $\partial_\rho^2 g_{ab}$ are $1/n$-suppressed and we get a first-order flow equation for the metric,
\begin{eqnarray}
\fr{\rho}\partial_\rho  g_{ab} = 2R_{ab}\,, \label{eq:ricci-flow-rho}
\end{eqnarray}
which, in terms of the parameter $\lambda$ in \eqref{eq:taurho}, takes the conventional form of the Ricci flow equation \eqref{eq:ricci-flow-tau}.

The equation for $K^i{}_j$ determines $N_1$,
\begin{eqnarray}
N_1= \frac{\rho}{4}\partial_\rho \ln g = \fr{2}\rho^2 R,
\end{eqnarray}
where in the second equality we have used (\ref{eq:ricci-flow-rho}).
With these equations, the scalar and vector constraints are automatically satisfied up to the relevant order.

Thus, solutions of the Ricci flow \eqref{eq:ricci-flow-rho} provide complete Einstein vacuum metrics \eqref{eq:riccflmetric} to leading order in $1/n$.

%
\section{Large-$D$ conifolds}\label{sec:2dmerger}

The Ricci flow equation simplifies considerably for the $p=2$ flows that are relevant to the black hole/black string transition in $D+3$ dimensions. This will allow us to find solutions which extend the horizon beyond the conical region and are asymptotically flat in the direction away from the horizon.

\subsection{Ricci flow as logarithmic diffusion}\label{sec:logdif}

We are interested in static black hole/black string configurations, which correspond to two-dimensional flows with one isometry. The two-dimensional metric can be written as
\beq
g_{ab}(\rho,y)dy^a dy^b =V(\rho,\xi)\lp -dt^2+ d\xi^2\rp\,.
\eeq
This part of the geometry, which has extent $1/D$ (in horizon radius units), can be regarded as describing the near-horizon region in the merger spacetime.

Then the Ricci flow equation \eqref{eq:ricci-flow-rho} becomes the partial differential equation
\begin{eqnarray}
- \fr{\rho} \partial_\rho V = \partial_\xi^2 \ln V\,. \label{eq:einstein-LO}
\end{eqnarray}
In terms of the flow parameter $\lambda$ in \eqref{eq:taurho} we find the one-dimensional logarithmic diffusion equation
\begin{eqnarray}\label{eq:logdiff}
 \partial_\lambda V = \partial_\xi^2 \ln V.
\end{eqnarray}
This is invariant under scaling
\begin{eqnarray}
 V(\lambda,\xi) \to \alpha^{-1} \beta^2 V(\alpha \lambda,\beta \xi),\label{eq:log-dif-scale}
\end{eqnarray}
and also under translations $\lambda \to \lambda + C$.

The complete spacetime metric \eqref{eq:riccflmetric} is of the form 
\begin{eqnarray}\label{eq:logdiffmetric}
 ds^2 = \left(1+\frac1{n}\frac{\rho \partial_\rho V}{V}\right)d\rho^2 + \frac{V(\rho,\xi)}{n}\left(-dt^2+d\xi^2\right)+\rho^2 d\Omega_n^2\,.
\end{eqnarray}
The horizon lies at $\xi=\xi_h$ such that 
\beq
V(\rho,\xi_h)=0.
\eeq
We see that $\rho$ is the coordinate along the horizon while $\xi$ is a coordinate in the direction away from the horizon.

The large-$D$ smoothed cones \eqref{eq:largeDcones} are recovered as the solution
\begin{eqnarray}
 V(\rho,\xi) = \frac{\rho^2-\mu}{\cosh^2\xi} \label{eq:largeDcone-V}\,,
\end{eqnarray}
once we transform the coordinates
\beq
\tan\chi=\sinh\xi\,.
\eeq

Observe also that \eqref{eq:logdiffmetric} admits flat space (in cylindrical coordinates) as the solution $V=1$. Solutions with the asymptotic behavior $V\to 1$ will then be asymptotically flat.

\subsection{Single black hole solution}\label{sec:single-BH}

Other solutions of the logarithmic diffusion equation are known in the literature which admit an interpretation in terms of large-$D$ black holes. A simple one is \cite{King93,Vez96}\footnote{This was originally suggested by de Gennes in his work on the spreading of microscopic droplets~\cite{deGennes1984}.}
\begin{eqnarray}
 V(\rho,\xi)= \fr{1+e^{-2(\xi+ \rho^2)}},\label{eq:sol-exp-type}
\end{eqnarray}
with
\beq
N(\rho,\xi)=1+\frac1{n}\frac{2\rho^2}{1+e^{2(\xi+\rho^2)}}\,.
\eeq
Since $V\to 1$ at $\xi\to \infty$ and at $\rho \to \infty$, the resulting metric is asymptotically flat in these limits. It has a horizon at $\xi\to-\infty$ with $\rho < \infty$.

This solution does not contain as a limit any of the previous ones, \eqref{eq:largeDcone-V}, since those are symmetric under $\xi\to -\xi$ and thus contain two different regions in which the horizon grows as $\rho\to\infty$. Instead, the solution \eqref{eq:sol-exp-type} is identified as a single Schwarzschild black hole in the large-$D$ limit \cite{Emparan:2013xia}. More precisely, it describes a portion of the spherical horizon near one of its poles, with the polar angle extending over a range $\ord{1/\sqrt{n}}$. We may think of it as related to the split conifolds (as we will confirm in the next section), but consisting of only one of the two horizons. In this case the Euclidean geometry does not have a topological $S^2$ but rather the familiar cigar shape of Euclideanized black holes. 

To make this connection, we change coordinates $(t,\xi,\rho)\to (\tilde{t},x,\theta)$ in the form
\beqa
t&=& \frac{n\, \tilde{t}\,}{2},\\
\xi&=&\ln\sinh(x/2)-\frac{n}{4}\sin^2\theta\,,\\
\rho&=&\frac{\sqrt{n}}{2}\sin\theta\lp 1+\frac{2}{n}\ln \cosh(x/2) \rp\,.\label{eq:rhotheta}
\eeqa
The metric then becomes
\begin{eqnarray}
 ds^2 = \frac{n}{4} \left[- \tanh^2(x/2)d\tilde{t}^2+\frac{dx^2}{n^2}+\left(1+\frac{4}{n}\ln \cosh(x/2)\right) d\Omega_{n+1}^2 \right],\label{eq:largeDSchw}
\end{eqnarray}
where the $S^{n+1}$ appears from
\beq
d\Omega_{n+1}^2 =d\theta^2+\sin^2\theta\, d\Omega_{n}^2\,.
\eeq
We recognize the geometry \eqref{eq:largeDSchw} as the large-$D$ near-horizon limit of the Schwarzschild black hole, with the horizon at $x=0$ \cite{Emparan:2013xia,Emparan:2015hwa}.

The appearance of this solution in this analysis deserves further comment. First, observe that the entire metric \eqref{eq:largeDSchw} is multiplied by a factor $n$, which means that, relative to the conventional large-$D$ limit of the Schwarzschild black hole in \cite{Emparan:2013xia}, here the lengths are scaled up by $\sqrt{n}$. This blow-up of the geometry is necessary in order that the near-horizon region of the black hole, with radial extent $\sim 1/n$, is appropriately contained within the regions of radial size $\sim 1/\sqrt{n}$ considered in this paper.

Second, as a consequence of this blow-up, only a `cap' of polar extent $\theta=\ord{1/\sqrt{n}}$ of the sphere $S^{n+1}$ is captured. This is evidenced by the factor $\sqrt{n}$ in the definition of $\theta$ in \eqref{eq:rhotheta}. Notice that $\rho\to\infty$ with $\rho^2+\xi$ finite\footnote{These are surfaces of constant $V$, \ie constant gravitational potential.} corresponds to the limiting extent of the cap, where $\sqrt{n}\,\theta\to\infty$.

Finally, it is interesting to see how the solution \eqref{eq:sol-exp-type} captures the gravitational potential that the black hole creates away from its horizon. Consider a massive particle at the origin of flat space,
\beq
ds^2=-dt^2+dz^2+dr^2+r^2d\Omega_n^2\,.
\eeq
It generates a Newtonian potential
\beq
\Phi=\frac{r_0^n}{(r^2+z^2)^{n/2}}\,,
\eeq
where $r_0$ is the mass length-scale ---the horizon radius in the general-relativistic solution. Let us zoom in on a region of radius $\sim r_0$, of radial extent $\sim 1/n$ and within a distance $\Delta r \sim 1/\sqrt{n}$ of the $z$ axis at $r=0$. To this purpose, change coordinates $(z,r)\to (\xi,\rho)$,
\beqa
z&=&r_0\lp 1+\frac{2\xi}{n}\rp\,,\\
r&=&\frac{2r_0}{\sqrt{n}}\rho\,.
\eeqa
Then, when $n\to\infty$ the flat geometry is
\begin{eqnarray}
 ds^2 = \frac{4r_0^2}{n} \lp d\rho^2 + \fr{n}\lp-d\tilde{t}^2+d\xi^2\rp+ \rho^2d\Omega_n^2\rp,\label{eq:flatref}
\end{eqnarray}
where $\tilde{t} = n t/(2r_0)$, and the gravitational potential in this region becomes
\beq
\Phi\simeq e^{-2(\xi+\rho^2)}\,.
\eeq
This is the same as we obtain by linearizing the solution \eqref{eq:sol-exp-type} in \eqref{eq:logdiffmetric}.

\subsection{Topology-changing conifolds}\label{sec:double-horizon}

In our study of the logarithmic diffusion equation \eqref{eq:einstein-LO}, the King-Rosenau solution \cite{KingRosenau}
\begin{eqnarray}
V &=& \frac12\lp \tanh\lp\rho^2-\mu+\xi\rp+\tanh\lp\rho^2-\mu-\xi\rp\rp\nn\\
&=&\frac{\sinh( 2 (\rho^2-\mu))}{\cosh( 2 (\rho^2-\mu))+\cosh(2 \xi)}\label{eq:KR-sol}
\end{eqnarray}
plays a very important role. We will refer to it as the KR conifold.
It gives
\begin{eqnarray}
N =1+\frac1{n}\frac{2\rho^2}{\sinh\lp 2 (\rho^2-\mu)\rp}\left(\frac{1+\cosh\lp 2 (\rho^2-\mu)\rp \cosh(2\xi)}{\cosh\lp 2 (\rho^2-\mu)\rp +\cosh(2\xi)}\right).\label{eq:KR-sol-B}
\end{eqnarray}

In contrast to the previous single-black hole solution \eqref{eq:sol-exp-type}, this one is symmetric in $\xi\to-\xi$ and therefore the horizon is present on two sides, at $\xi \to \pm \infty$, for all the range of finite $\rho$.
Since $V\to 1$ in the limit $\rho\to \infty$ at finite $\xi$, the solution is asymptotically flat in that direction. 

Consider \eqref{eq:KR-sol} at large distance $\rho\gg \sqrt{|\mu|}$ (or $\rho\gg 1$ if $\mu=0$) and close to the horizon at $\xi\to -\infty$, keeping $\xi+\rho^2$ finite. We find
\beq
V=1-e^{-2(\xi+\rho^2)}+\ord{e^{-4\rho^2},e^{4\xi}}\,,\label{eq:KR-oneside-as} 
\eeq
which is the same asymptotic behavior close to the horizon as in the single black hole solution \eqref{eq:sol-exp-type}. By symmetry, we get a mirror of this at $\xi\to\infty$. Thus, using the same coordinate change~(\ref{eq:rhotheta}), we find that the horizon geometry in each of the two directions away from the conical region approaches a spherical cap.

The large-$D$ cones that we found earlier in \eqref{eq:largeDcones} and \eqref{eq:largeDcone-V} can be recovered in a scaling limit of the KR conifold.
We use \eqref{eq:log-dif-scale} with 
\beq
\lambda = \frac{\mu-\rho^2}2,
\eeq
in order to take the limit $\lambda\to 0$ as
\begin{eqnarray}
\lim_{\alpha \to 0}  \frac{1}{\alpha}V(\alpha \lambda,\xi) =  -\frac{2 \lambda}{\cosh^2\xi}=\frac{\rho^2-\mu}{\cosh^2\xi}.\label{eq:KRscaling}
\end{eqnarray}

The KR conifolds \eqref{eq:KR-sol} have the same topology as the smoothed cones \eqref{eq:largeDcone-V} with the same sign of $\mu$. Therefore, they connect the topology-changing region to an asymptotically flat region, and can be appropriately called fused, critical and split conifolds.

Setting $\xi=-\infty$ in \eqref{eq:KR-sol} we obtain the spatial geometry of the horizon,
\begin{eqnarray}
ds^2_H = \left(1+\frac{4 \rho^2}{n} \coth(2(\rho^2-\mu))\right)d\rho^2 + \rho^2 d\Omega_n^2\,.\label{eq:horizonds2}
\end{eqnarray}
This geometry across the topology-changing transition $\mu>0\leftrightarrow \mu<0$ can be adequately illustrated by embedding it into the auxiliary flat space
\begin{eqnarray}
 ds_H^2=\fr{n}dX^2+dY^2+Y^2 d\Omega_n^2.\label{eq:flatembed}
\end{eqnarray}
The embeddings are presented in figure \ref{fig:embedding}. The details can be found in appendix~\ref{app:embedding}.

The KR conifold breaks the scale invariance of the exact cone both at short and long distances. The short-distance breaking is controlled by the parameter $\mu$, while at large distances there must appear a length scale related to the only length parameter of the black hole/black string system near the critical point, namely $r_0$, the maximum horizon thickness of the $S^n$.\footnote{In the large $n$ limit the scale of the solution could also be given by, say, $r_0/n$ or $r_0/\sqrt{n}$, which are parametrically different than $r_0$. But, actually, it is the latter that sets the size in this instance.} However, no such length parameter is visible in the conifold geometry ---it must have been set to one. We will see in section \ref{sec:doublehole-interaction} that this length scale is $2r_0$, or equivalently $L$, the length of the compact Kaluza-Klein circle. Thus, dimensions can be restored in the metric \eqref{eq:logdiffmetric} by multiplying it by $(2r_0)^2$.

A subtlety arises in that the limit $\lambda\to 0$ defined in \eqref{eq:KRscaling} requires in general a scaling of the parameter $\mu$, \ie it is not a limit of a given solution but a limit along a family of solutions. Therefore we cannot immediately conclude that a KR solution \eqref{eq:KR-sol} with a given $\mu$ contains a small region with the conifold geometry \eqref{eq:largeDcones} of that $\mu$. Such a region exists when we can take $\lambda\to 0$ independently of the size of $\mu$, which is possible when $\mu\geq 0$. In this case, we can zoom in at $\rho\simeq \sqrt{\mu}$, to find\footnote{In appendix~\ref{app:coneperturb} we show that the leading corrections to \eqref{eq:merger-waist} with $\mu=0$ correctly reproduce, in the large-$D$ limit, the perturbations of the double cone studied in \cite{Asnin:2006ip}.}
\begin{eqnarray}
 V \simeq \frac{\rho^2-\mu}{\cosh^2\xi}+\Ord{(\rho^2-\mu)^3}\,.
\label{eq:merger-waist}
\end{eqnarray}

This limit is not possible in solutions with  $\mu<0$, so the split cones of \eqref{eq:largeDcones} appear as small regions of split KR solutions only when $|\mu|$ decreases along with $\rho^2$. This, however, is not a problem for the interpretation that we are giving of these solutions. Indeed, next we proceed to show that they describe a two-black hole geometry, in a similar way that \eqref{eq:sol-exp-type} corresponds to a single black hole.

\subsection{Interaction between two black holes}\label{sec:doublehole-interaction}

Consider the KR solution \eqref{eq:KR-sol} with large negative $\mu\to -\infty$. Near the horizon at each of the two sides, $|\xi|\gg 1$, we recover the single black hole solution \eqref{eq:sol-exp-type} (after appropriately shifting $\rho^2\to\rho^2+\mu/2$ and $\xi\to \xi-\mu/2$),
\beq
V\to \frac{1}{1+e^{-2(-|\xi|+\rho^2)}}\,.
\eeq
As we will see, this is consistent with the fact that $-\mu$ measures the separation between the two black holes.

Remarkably, the KR solution also accounts for the gravitational attraction between the two black holes when their separation is finite ---and even when they overlap, as in the case of the fused conifold with $\mu>0$.

To see this, take the gravitational potential from two identical sources,
\begin{eqnarray}
\Phi = \frac{r_0^n}{\lp r^2+z^2\rp^{n/2}}+\frac{r_0^n}{\lp r^2+(z-L)^2\rp^{n/2}},
\end{eqnarray}
where $L$ denotes the distance between them ---in our case, the length of the compactified circle in the $z$ direction.\footnote{Out of the infinite set of images of the black hole in the circle, only the closest one has a significant attractive effect when $n\to\infty$.}
We assume that the two black holes almost touch each other,
\begin{eqnarray}
 L = 2 r_0+\ord{1/n}.
\end{eqnarray}
As we did for the single black hole, we zoom in on this region with a radial extent $\sim 1/n$ and a distance $\sim 1/\sqrt{n}$ off the axis, by changing coordinates
\begin{eqnarray}
r = \frac{L\rho}{\sqrt{n}},\quad z = \frac{L}{2} \left(1+\frac{2\xi}{n}\right). \label{eq:near-polar-coordinate}
\end{eqnarray}
Then, in the limit $n\to\infty$ the potential results in
\begin{eqnarray}
 \Phi \simeq  2 e^{-2(\rho^2-\mu)}\cosh(2\xi), \label{eq:double-src}
\end{eqnarray}
where $\mu$ is defined by
\begin{eqnarray}
 e^{2\mu} = \left(\frac{2r_0}{L}\right)^n. \label{eq: close-horizon-assumption }
\end{eqnarray}
This reproduces the expansion of the solution \eqref{eq:KR-sol} in this region with the correct value of $\mu$,\footnote{Remarkably, one can easily guess the complete solution~\eqref{eq:KR-sol} from knowledge of the linearized result \eqref{eq:double-src}.
Assuming an expansion of the form
\begin{eqnarray}
V(\rho,\xi) = 1- \sum_{j=1}^\infty v_j(\xi) e^{-2j(\rho^2+\mu)}
\end{eqnarray}
with $v_1(\xi) = 2\cosh(2\xi)$ and inserting it in the logarithmic diffusion equation~\eqref{eq:einstein-LO}
gives
\begin{eqnarray}
 V(\rho,\xi) = 1+ 2\sum_{j=1}^\infty (-1)^j\cosh(2j\xi) e^{-2j(\rho^2+\mu)}.
\end{eqnarray}
This is then resummed to the closed form of (\ref{eq:KR-sol}).
 }
and it gives the precise relation between the parameter $\mu$ in the topology-changing solution and the gap between the two horizons,
\begin{eqnarray}\label{eq:gap}
 \Delta z=L -2r_0 \simeq -2L\frac{\mu}{n}.
\end{eqnarray}

\begin{figure}[t]
\begin{center}
\includegraphics[width=\linewidth]{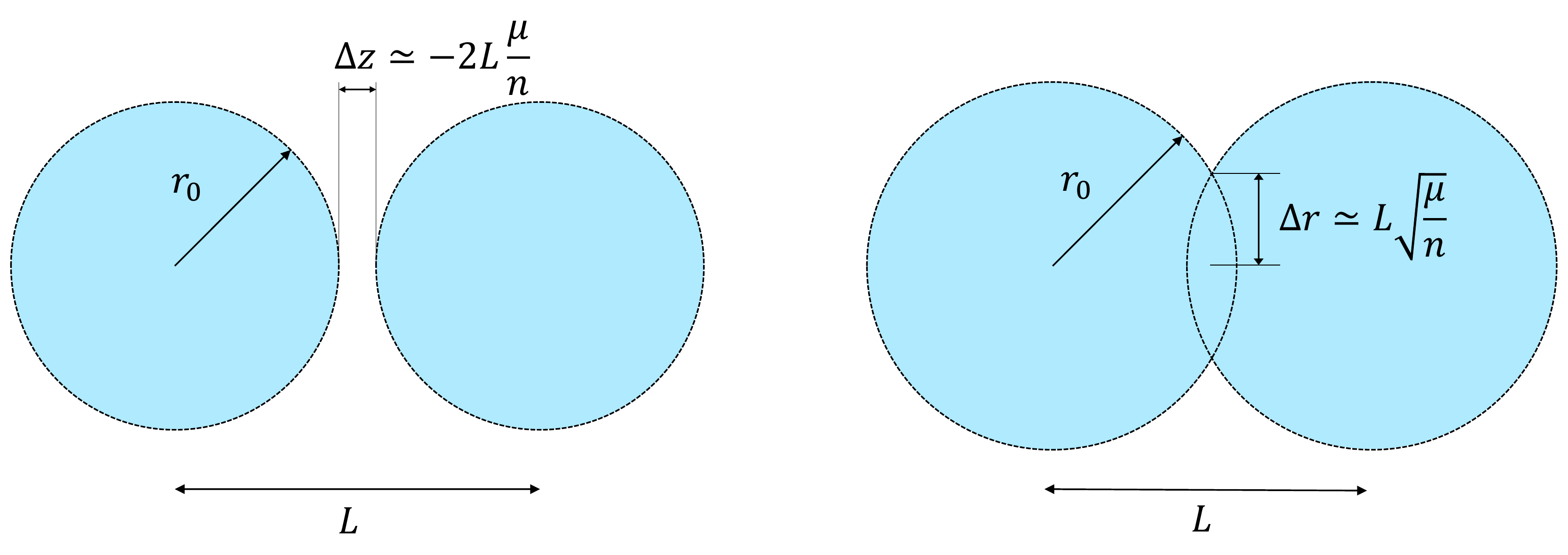}
\caption{The Newtonian potential for a double point source gives the parameters of the KR conifold solution, both in split ($\mu<0$, left) and in fused ($\mu>0$, right) configurations.}
\label{fig:doublebh}
\end{center}
\end{figure}

This analysis is actually also valid when $\mu>0$, in which case $r_0>L/2$ and $\Delta z<0$, and we would have the two black holes overlap. Then we can easily find the distance from the axis at which the two horizons would cross,
\beq
\Delta r\simeq \sqrt{r_0|\Delta z|}\simeq L\sqrt{\frac{\mu}{n}}\,,
\eeq
which gives us the thickness of the neck in fused conifolds with the correct scaling. We illustrate this in figure \ref{fig:doublebh}.

It is in fact possible to do a little better and integrate the distance in the KR split conifold along the axis $\rho=0$ from $\xi=-\infty$ to $\xi=\infty$, to obtain the exact separation (at leading large-$D$ order) between the horizons,
\beq
\Delta z= \frac{2L}{n}\sqrt{\tanh|\mu|}\, K\lp\tanh^2|\mu|\rp \label{eq:delzexact}
\eeq
(see \eqref{eq:delzint}). Here $K(x)$ is the complete elliptic integral of the first kind and we have taken into account that the conifold metric \eqref{eq:riccflmetric} must be multiplied by a prefactor $L^2/n$ in order to obtain the physical metric (cf.~\eqref{eq:flatref}). In the large separation limit $\mu \to -\infty$ this result reproduces the gap \eqref{eq:gap}. At small $|\mu|$, instead, we find
\beq
\Delta z \to \pi L\frac{\sqrt{|\mu|}}{n}\,,
\eeq
which can be directly obtained from the split cone \eqref{eq:largeDcones}. This is a smaller distance than the extrapolation of \eqref{eq:gap} to small $|\mu|$, which reflects the effect that at very short separations the attraction distorts the horizons in the region near the axis and brings them more closely together. In fact, the remarkable fact that the simple KR conifold solution accounts for the full non-linear interactions of General Relativity in these configurations is made manifest in the non-linear dependence of $\Delta z$ on $|\mu|$ in \eqref{eq:delzexact}.

\medskip

We have now derived a detailed picture of the black hole/black string configurations near the merger in the region close to the polar axis. They are well approximated as an array of almost spherical blacks holes which are either barely separated, or overlapping into a highly non-uniform black string. We will next see that far, from the polar axis, the conifold geometry is indeed, both for $\mu>0$ and $\mu<0$, that of slightly deformed Schwarzschild black holes.

\subsection{Beyond the polar cap: deformation of the Schwarzschild black hole}

In  section \ref{sec:single-BH}, we observed that the single black solution is exactly identified as the polar cap of the large-$D$ Schwarzschild black hole, extending over a range of polar angles $\sim 1/\sqrt{n}$. 

We now show that the KR conifold, for positive, negative or zero $\mu$, also matches the same Schwarzschild geometry sufficiently away from the conical region, on each of the two sides of the horizon, $\xi\to\pm \infty$, and then study how this large-$D$ black hole is slightly deformed in the conifold configuration. For definiteness, we choose to study the horizon at $\xi=-\infty$, with the analysis for $\xi\to\infty$ being essentially identical by symmetry.

At $\rho\to\infty$ and close to the horizon $\xi\to-\infty$, the KR conifold has the same equipotential surface~\eqref{eq:KR-oneside-as} as the single black hole solution. Therefore, $\rho$ in (\ref{eq:rhotheta}) is again an appropriate coordinate along the horizon at $\xi=-\infty$.
As discussed in section \ref{sec:single-BH}, the limiting extent of the cap, where it matches to the complete Schwarzschild geometry, lies at $\sqrt{n}\theta\to \infty$.
More precisely, the matching condition is\footnote{The stronger condition $1\ll n\sin^2\theta \ll n$ yields the same matching result.}
\begin{equation}
1 \ll e^{n\sin^2\theta} \ll e^n. \label{eq:matching-condition}
\end{equation}
where $\theta$ is the polar angle introduced in \eqref{eq:rhotheta}.
Under this condition, the KR metric is expanded as
\begin{align}\label{eq:KR-expand-1st-metric}
ds^2 = \frac{n}{4} &\left[ (1+\delta A)  \left(-\tanh^2(x/2) d\tilde{t}^2 +\frac{dx^2}{n^2}\right)\right. \nonum
&\left. + \left(1+\frac{4}{n} \ln \cosh(x/2)\right)((1+\delta B) d\theta^2+d\Omega_n^2)+ 2\delta C d\theta dx \right]+\ord{e^{-2n\sin^2\theta}},
\end{align}
where $\tilde{t}=2t/n$ and the leading corrections are
\begin{subequations}\label{eq:KR-expand-1st}
\begin{align}
&\delta A = -e^{4\mu} \cosh^2(x/2) e^{-n\sin^2\theta} ,\\
&\delta B= 2 e^{4\mu} \sin^2\theta e^{-n\sin^2\theta},\\
&\delta C=\fr{2} e^{4\mu} \sinh(x) \sin\theta e^{-n\sin^2\theta}.
\end{align}
\end{subequations}
Taking the limit $e^{-n\sin^2\theta} \to 0$, the KR conifold reproduces the large-$D$ Schwarzschild solution (\ref{eq:largeDSchw}). This means that 
the inner conical region at smaller $\rho$, which is determined by the parameter $\mu$, does not make any difference in the leading-order matching.\footnote{Here we only refer to the near horizon geometry. As seen in the previous section, the value of $\mu$ is directly related to the separation between the two black holes.}
The reason is that the effect of the interaction between the two black holes quickly decays at polar angles $\theta >1/\sqrt{n}$ outside the conifold region as $e^{-n\sin^2\theta}$. This is a manifestation of the generic exponential-in-$D$ localization of the gravitational interaction when $D$ is large \cite{Emparan:2013moa}.

The information about the inner conical geometry (and topology) is, then, carried by the correction terms in \eqref{eq:KR-expand-1st}. Let us see that these corrections are actually identified as scalar perturbations of the Schwarzschild black hole (we defer some of the details to appendix~\ref{app:schwpert}).

For geometries with $S^{n+1}$ symmetry, such as the leading order metric~\eqref{eq:largeDSchw}, any perturbation can be decomposed in spherical harmonics,
\beq
\Pi_\ell(\theta) = C^\frac{n}{2}_\ell(\cos\theta)\,,
\eeq
where $C^\frac{n}{2}_\ell(x)$ are Gegenbauer polynomials. Using the results in \eqref{eq:s-harmonics-largeD}, the leading corrections are scalar harmonics with $\ell=n$,
\begin{eqnarray}
e^{-n\sin^2\theta} \simeq \sigma_n \Pi_n (\theta), \label{eq:match-Y}
\end{eqnarray}
where the coefficient $\sigma_n$ is given by the value at the pole $\theta=0$,
\begin{eqnarray}
 \sigma_n =\fr{C^\frac{n}{2}_n(1)}\simeq \frac{\sqrt{\pi n} }{2^{2n-1}}.\label{eq:sigma-value}
\end{eqnarray}
These corrections to the geometry can be regarded, in the terminology of \cite{Emparan:2014aba}, as `non-decoupled', being localized at the boundary of the near-horizon region and having very large wavenumber $\sim D$.
As such, their amplitude is $\ord{e^{-D}}$,  which is non-perturbatively small in the $1/D$ expansion and invisible in the large $D$ effective theory.

In order to verify that the KR conifold corrections \eqref{eq:KR-expand-1st} match to the scalar harmonic perturbations, we write the latter as
\begin{align}
ds^2 =& \frac{n}{4}\left[- \tanh^2(x/2)(1+\delta h^t{}_t) d\tilde{t}^2 + (1+\delta h^x{}_x) \frac{dx^2}{n^2}\right.\nonum
&\left.+ \left(1+\frac{4}{n}\ln \cosh(x/2) \right) (\gamma_{IJ}+\delta h_{IJ}) dz^I dz^J+2\delta h_{xI} dx dz^I \right]\label{eq:KR-match-BH-pert}
\end{align}
where $\gamma_{IJ}$ is the metric for $S^{n+1}$, and
\begin{subequations}
\begin{align}
&\delta h^t{}_t  =\delta  h^x{}_x = - \sigma_n e^{4\mu}  \cosh^2(x/2) \Pi_n(\theta)\,,\\
&\delta h_{xI} = -\frac{\sigma_n e^{4\mu}}{4n} \sinh(x) {\sf V}_I(\theta)\,, \\
&\delta h_{IJ} = \frac{\sigma_n e^{4\mu} }{2n^2} {\sf T}_{IJ}(\theta)\,.
\end{align}
\end{subequations}
We see that indeed the purely scalar perturbations $\delta h^t{}_t$ and $\delta  h^x{}_x$ are correctly reproduced.
The ${\sf V}_I(\theta)$ and ${\sf T}_{IJ}(\theta)$ are the scalar-derived vectors and tensors made from $\Pi_n(\theta)$. The details of the matching of the corresponding perturbations are given in appendix~\ref{sec:sn-harmonics-VT}.

The conclusion of this analysis is that, as we move further away from the cone region of the conifold, the geometry can be apropriately treated as a perturbation of the Schwarzschild black hole. Thus, our construction gives the entire geometry of the localized black hole and non-uniform black string phases in solution space near the merger transtion.

\subsection{Rigidity of compact ancient flows}

To conclude this long section, we mention that the KR conifold solution is called a two-dimensional compact ancient solution, since the geometry $g_{ab}$ that flows with $\lambda$ is a Euclidean, compact topological $S^2$. The single black hole solution is a non-compact ancient solution, since in this case the Euclidean geometry that flows is cigar-shaped with non-compact topology $\mathbf{R}^2$ ---indeed the solution is known as the {\it cigar soliton}.

It has been proven that the two-dimensional compact ancient solution should be either one of the cones \eqref{eq:largeDcone-V}, with exact $SO(3)$ symmetry, or one of the
KR solutions \eqref{eq:KR-sol} \cite{Daskalopoulos+2012}.
This rigidity theorem strongly restricts the asymptotic behavior of the geometry at large $\rho$ in the mergers, as long as they can be regarded as flows of a two-dimensional geometry.

The KR conifold can also be described using concepts of statistical mechanics and field theory.\footnote{We thank our referee for suggesting this.} Namely, the scale-breaking induced by the parameter $\mu$ can be regarded as a relevant perturbation of the exact cone at short distances, while the critical KR cone (with $\mu=0$) is a relevant perturbation of the cone at long distances. The solution is therefore an attractor.

%
\section{Neck resolution in fused conifolds}\label{sec:neck-resolve}

In the previous sections we have shown how the large-$D$ expansion yields a complete resolution of the singular double cone into a regular split cone, eq.~\eqref{eq:largeDcones} with $\mu<0$. 
We have also obtained a fused cone geometry, \eqref{eq:largeDcones} with $\mu>0$, for which the KR conifold provides an extension away from the conical region. However, as we noted in section \ref{sec:cones}, these fused geometries are still singular at the neck $\rho=\sqrt{\mu}$.

\begin{figure}[t]
\begin{center}
\includegraphics[width=.35\linewidth]{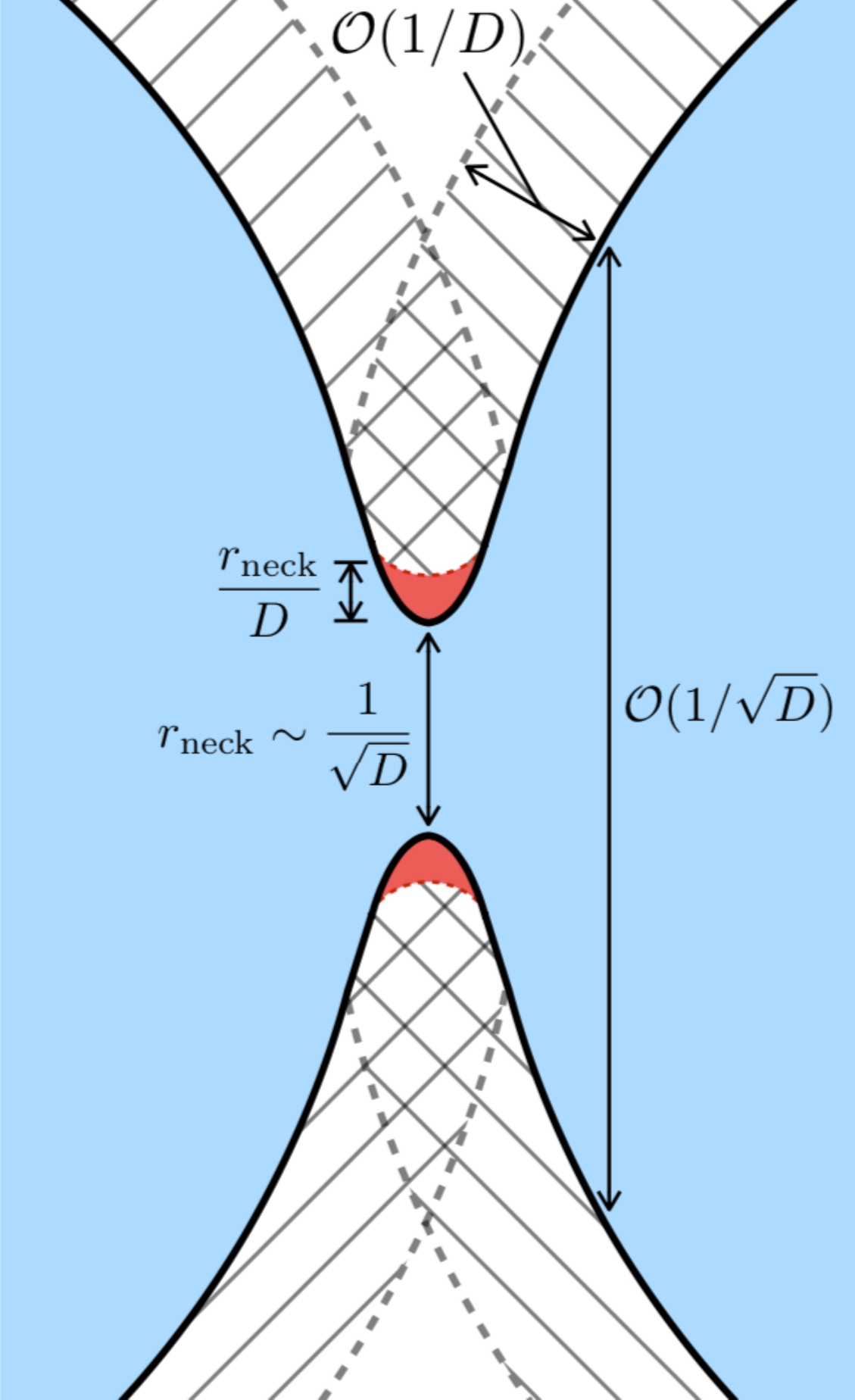}
\caption{The different regions of the neck in the fused conifold, and their sizes. Blue: non-uniform black string. Hatched: near-horizon region, of size $\ord{1/D}$. Red: near-neck region, of size $\ord{1/D^{3/2}}$. The latter is the region resolved in section \ref{sec:neck-resolve}.}
\label{fig:neckresolution}
\end{center}
\end{figure}

This singularity is the result of not enough resolution of the neck in the large-$D$ expansion that we have performed. The radius of the neck has size $\ord{1/\sqrt{D}}$, and its near horizon region will be smaller by a factor $1/D$, which is too small to resolve in our previous conifold constructions. We illustrate these features in figure \ref{fig:neckresolution}.
%
We will see next that the singularity of the fused cones is resolved by adequately blowing up the neck region.\footnote{Throughout this section, we only consider the fused conifold for black strings, $p=2$, but we expect that the results will be qualitatively the same for all $p\geq 2$.}

\subsection{Near-neck limit}\label{subsec:nearneck}
Let us first argue that the $1/n$ expansion that we have performed breaks down near the neck, requiring a different scaling to resolve it.
In the conifold ansatz \eqref{eq:largeDcone-ansatz} we introduce $1/n$ corrections as
\begin{eqnarray}
S(\rho)=(\rho^2-\mu)\left(1+\sum_{k=1}\fr{n^k} S_k(\rho)\right),\quad
 N(\rho)=1+\sum_{k=1} \fr{n^k} N_k(\rho).
\end{eqnarray}
Solving the equations up to the next-to-leading order, we obtain
\begin{subequations}\label{eq:sol-neck-nlo}
\begin{eqnarray}
 S = (\rho^2-\mu)\left(1+\fr{n}\frac{2\ln(\rho^2/\mu-1)+\rho^2/\mu-1}{\rho^2/\mu-1}\right),\label{eq:sol-neck-nlo-R}
\end{eqnarray}
and
\begin{eqnarray}
 N = 1+\frac{1}{2n}\frac{\rho^2}{\rho^2-\mu}.\label{eq:sol-neck-nlo-B}
\end{eqnarray}
\end{subequations}
An integration function in (\ref{eq:sol-neck-nlo-R}) has been absorbed by a change of $\mu$, which does not affect the physics. For concreteness, it is chosen to avoid the first order pole $1/(\rho^2-\mu)$ in each $S_k$.

It is obvious that the $1/n$ expansion in \eqref{eq:sol-neck-nlo} breaks down as one approaches the dangerous neck region $\rho\approx \sqrt{\mu}$. But we can also see that we should expect to match it to a finer {\it near-neck solution} obtained by performing a slightly different expansion.

Let us introduce a near-neck coordinate $\sigma$ by
\begin{eqnarray}\label{eq:nearneckrho}
 \rho =\sqrt{\mu} \left(1+\frac{\sigma}{n}\right)\,.
\end{eqnarray}
Then, when $n\to\infty$ the metric will take the form 
\begin{eqnarray}
 ds^2 \simeq \mu \left(\left(1+\ord{n^0}\right)\frac{d\sigma^2}{n^2}+\left(1+\ord{n^0}\right)\frac{2\sigma}{n^2}d\Omega_2^2+d\Omega_{n}^2\right),\label{eq:neck-sol-near-neck-limit}
\end{eqnarray}
where we assume that the sum of all higher corrections in $1/n$ remains finite. Actually, one can see that each $R_k$ has at most a $k$-th pole of $\rho^2-\mu$, which does give an $\ord{1}$ contribution. Note also that deformations that break the $SO(3)$ symmetry become negligible in the near-neck limit, since the perturbative analysis in \cite{Asnin:2006ip}  (see appendix~\ref{app:coneperturb}) implies that these deformations are suppressed by at least one power of $(\rho^2-\mu)^2$.

Observe also that the form of the near-neck metric (\ref{eq:neck-sol-near-neck-limit}) requires another new type of large-$D$ limit, in which there are \textit{two spatial directions} in which the gradients are large and of order $\ord{D}$ (we are not including in this count the isometric time direction from the Lorentzian Wick-rotation of the $S^2$).
This same scaling can be inferred from a perturbation analysis of the conical solution around $\rho=0$, which was performed in \cite{Kol:2002xz}: in appendix~\ref{app:pertneck} we show that if one requires that the perturbative-in-$\rho$ solution has a well-defined limit for $n\to\infty$, then the large-$D$ scaling above results.

\subsection{Near-neck solution}\label{sec:near-neck}
Following \eqref{eq:neck-sol-near-neck-limit}, we take a near-neck ansatz of the form
\begin{eqnarray}
 ds^2 =\mu\left(\frac{d\varsigma^2}{n^2}+\fr{n^2}\alpha^2(\varsigma)d\Omega_2^2+e^\frac{2\beta(\varsigma)}{n} d\Omega_{n}^2\right).
\label{eq:near-neck-ansatz}
\end{eqnarray}
Einstein's equations are
\begin{align}
&\alpha (\alpha''+\alpha'\beta')+\alpha'^2-1=0\label{eq:neckeq0-a},\\
&\frac{2 \alpha''}{\alpha}+\beta''+\frac{\beta'^2}{n}=0\label{eq:neckeq0-b},\\
&\beta''+\frac{2 \alpha'}{\alpha} \beta'+\beta'^2-e^{-\frac{2 \beta}{n}}=0.\label{eq:neckeq0-c}
\end{align}
The first equation reduces to
\begin{eqnarray}
 \beta' = \frac{1-\alpha'-\alpha\alpha''}{\alpha\alpha'}\label{eq:neckeq-dB}.
\end{eqnarray}
Taking the large-$n$ limit and eliminating $\beta'$ and $\beta''$ in (\ref{eq:neckeq0-c}), we obtain a single ODE,
\begin{eqnarray}
\alpha '^4-1+2 \alpha  \left(\alpha '^2+1\right) \alpha ''+\alpha ^2 \left(\alpha '^2-\alpha ''^2\right)=0. \label{eq:neckeq-A}
\end{eqnarray}
In the large-$D$ limit, (\ref{eq:neckeq0-b}) also gives $\beta''$ in terms of $\alpha$, whose integrability
with (\ref{eq:neckeq-dB}) is guaranteed by (\ref{eq:neckeq-A}).
Therefore, (\ref{eq:neckeq-A}) is our master equation for $\alpha$, and then $\beta$ is obtained from (\ref{eq:neckeq-dB}).

Assuming $\alpha(\varsigma) \to 0$ at $\varsigma \to 0$ with $\alpha'(0),\ \alpha''(0)\neq 0$, (\ref{eq:neckeq-A}) has a unique solution around $\varsigma=0$,
\begin{eqnarray}
 \alpha(\varsigma) = \varsigma \left(1-\frac{\varsigma^2}{36}+\frac{29\varsigma^4}{21600}+\ord{\varsigma^6}\right). \label{eq:neck-exp-sols-LD}
\end{eqnarray}
This coincides with the large-$D$ limit of the regular perturbative solution found in \cite{Kol:2002xz} (see (\ref{eq:neck-exp-sols})).

\begin{figure}[t]
\begin{center}
\includegraphics[width=8cm]{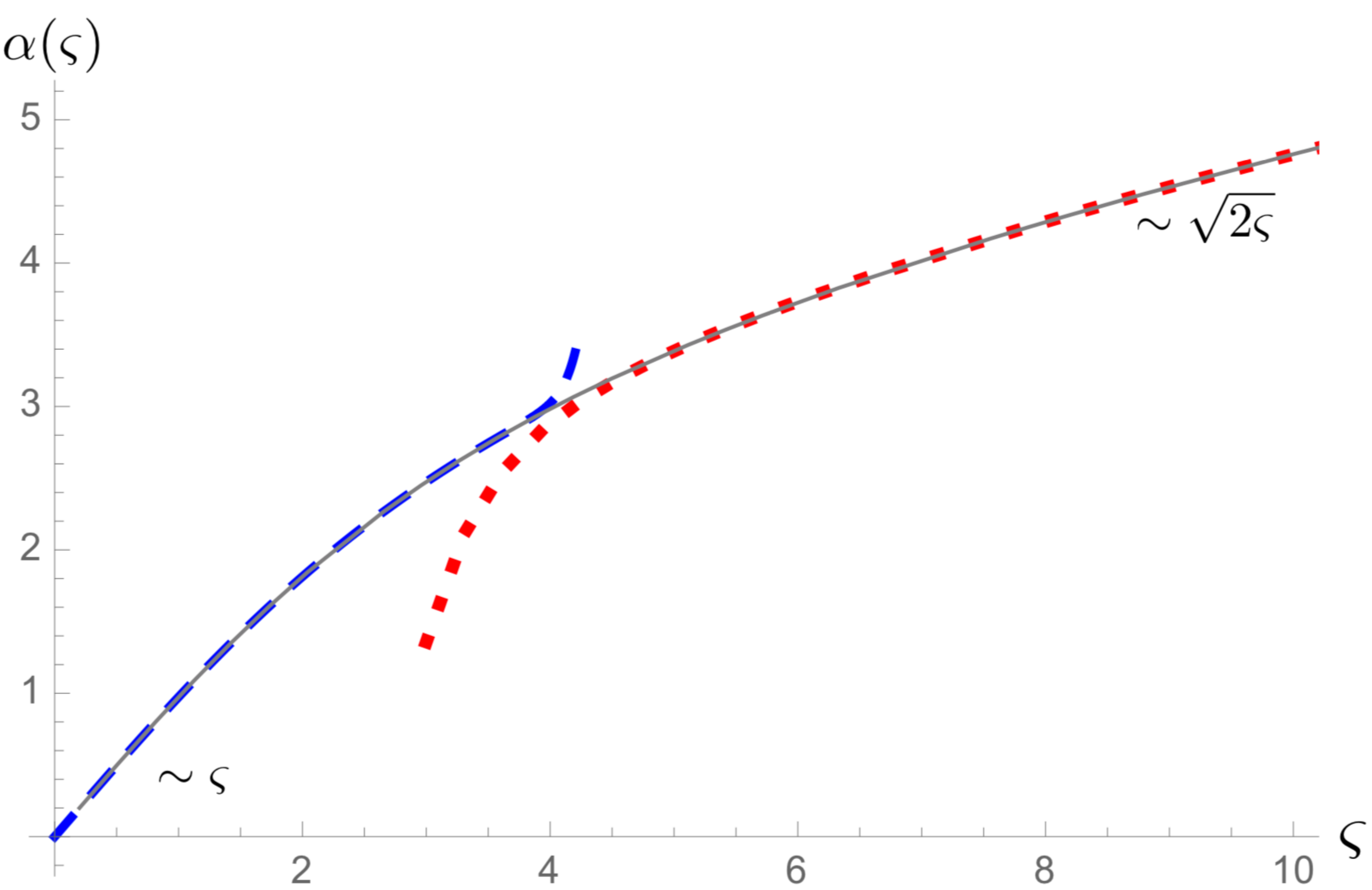}
\caption{Growth of $S^2$-radius in the near-neck region. The solid curve is the numerical result,
the blue dashed curve, the series expansion from the center~(\ref{eq:neck-exp-sols-LD}), which is also used for the initial condition for the numerics, the red dotted curve, the asymptotic expansion~(\ref{eq:neck-alpha-exp-as}) with the translation $\varsigma\to \varsigma+\Delta \varsigma$, where $\Delta \varsigma\simeq0.216$.}
\label{fig:nearnecksol}
\end{center}
\end{figure}

Numerical integration of (\ref{eq:neckeq-A}) from $\alpha(0)=0$ shows that, as expected, $\alpha$ increases monotonically (figure \ref{fig:nearnecksol}).
Actually, (\ref{eq:neckeq-A}) gives the asymptotic behavior $\alpha \simeq \sqrt{2\varsigma}$ at large $\varsigma$, which can be further expanded to find
\begin{eqnarray}
\alpha(\varsigma)  = \sqrt{2\varsigma}\left(1+\frac{\ln \varsigma}{4\varsigma}-\frac{(\ln \varsigma)^2-4\ln \varsigma+14}{32\varsigma^2}+\ord{\varsigma^{-3}}\right).
\label{eq:neck-alpha-exp-as}
\end{eqnarray}
At each order, the power of $\ln \varsigma$ is at most the same as the order in $1/\varsigma$.
In order to match this to the small $\varsigma$ expansion, we must shift $\varsigma\to \varsigma+\Delta \varsigma$, with a numerical fit of $\Delta \simeq 0.216$. The asymptotic expansion solution gives
\begin{eqnarray}
\beta(s) =\beta_0+\varsigma-\frac{\ln \varsigma}{2}-\frac{2\ln \varsigma-1}{8\varsigma}+\ord{\varsigma^{-2}}
\label{eq:neck-beta-exp-as}
\end{eqnarray}
where $\beta_0$ is an integration constant. The large-$D$ limit requires $\beta\ll n$, and therefore $\varsigma\ll n$.

To complete the analysis, in appendix~\ref{app:matchnearneck} we show that the near-neck limit of the fused conifold~(\ref{eq:sol-neck-nlo}) can be smoothly connected to the near-neck metric~(\ref{eq:near-neck-ansatz}).

\begin{figure}[t]
\begin{center}
\includegraphics[width=.8\linewidth]{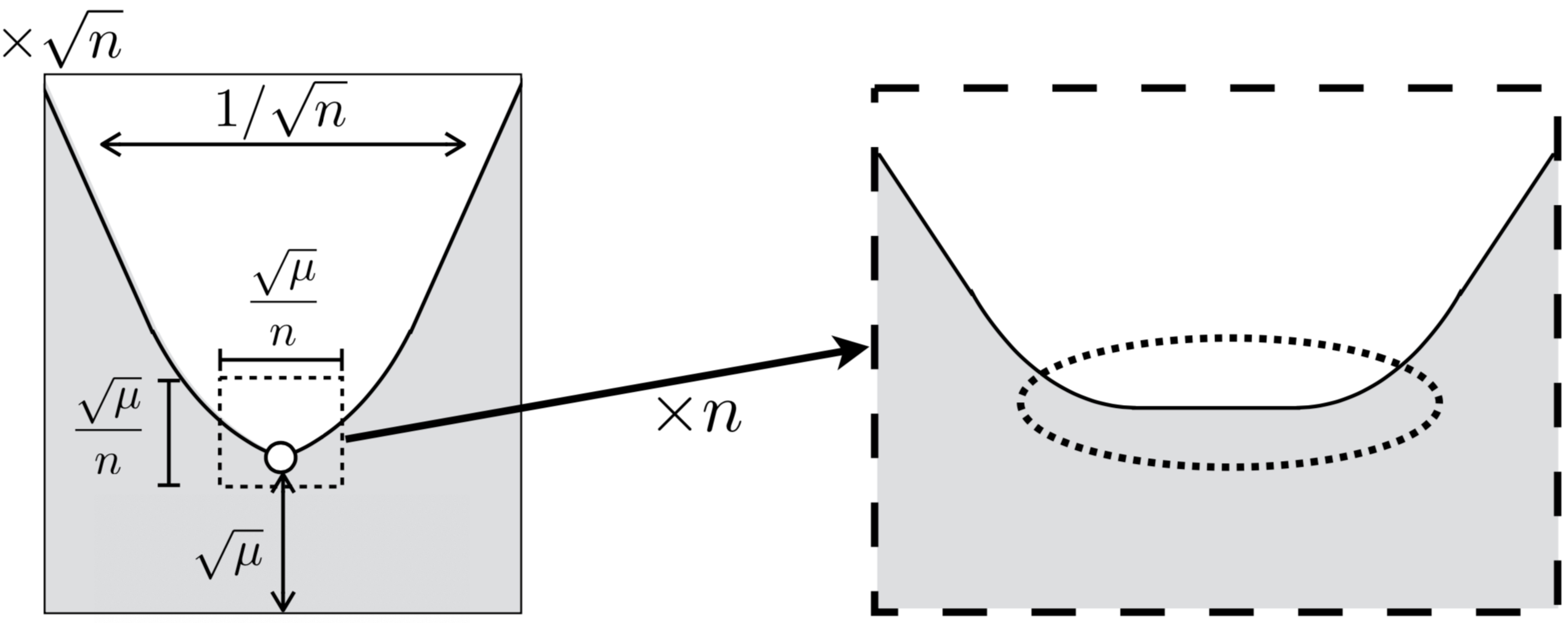}
\caption{Near-neck region. The left diagram is the neck region in the scaling of section \ref{sec:cones}. The right diagram is the blow-up of the near-neck region to obtain a smooth neck solution.}\label{fig:nearneckresolution}
\end{center}
\end{figure}

Figure \ref{fig:nearneckresolution} illustrates our result that the singular neck that appears in the fused cone at lengths of order $\ord{1/\sqrt{n}}$ is smoothly resolved by considering the near-neck substructure at scales $\ord{1/n}$.

\section{Extensions of Ricci flow mergers}\label{sec:extensions}

Although we will not consider any explicit solutions, it is possible to extend the Ricci flow equations of section \ref{sec:ricciflow} beyond the vacuum Einstein theory.

\subsection{Cosmological constant}

First we include a cosmological constant
\beq\label{eq:lambda}
\Lambda=-\frac{(n+p)(n+p-1)}{2L^2}\,,
\eeq
and look for solutions of the type 
\begin{eqnarray}
ds^2 =N^2(\rho,y) d\rho^2 + \fr{n}g_{ab}(\rho,y) dy^a dy^b+\rho^2 d\Sigma_{(k)n}^2\,,
\end{eqnarray}
where $d\Sigma_{(k)n}^2$ is the metric of a maximally symmetric space with curvature $k=0,\pm 1$. With a negative $\Lambda$, all of these are possible, but a zero or positive $\Lambda$ (with $L=i \hat{L}$) only allows the sphere $S^n$, with $k=1$. 

Retracing our steps of section \ref{sec:ricciflow}, as is done in appendix~\ref{app:cosmoflow}, we obtain the Ricci-flow-like equation
\begin{equation}
 \frac{k+\rho^2/L^2}{\rho} \partial_\rho g_{ab} = 2 R_{ab} + \frac{2}{L^2} g_{ab}. \label{eq:ricci-flow-L0}
\end{equation}

If we consider the conformally related geometry
\beq
\tilde{g}_{ab} = \fr{k+\rho^2/L^2} g_{ab},
\eeq
it satisfies the Ricci flow equation
\begin{equation}
\partial_\lambda \tilde{g}_{ab} = -2 \tilde{R}_{ab} \label{eq:ricci-flow-L1}
\end{equation}
with parameter
\begin{equation}
\lambda = \fr{2} \frac{L^2}{k+\rho^2/L^2}+\lambda_\infty. \label{eq:ricci-flow-L1-tau}
\end{equation}

Notice that in the limit of $k=1$ and $L\to\infty$, 
\begin{eqnarray}
\lambda  =  \left(\frac{L^2}{2}+\lambda_{\infty}\right) -\frac{\rho^2}{2} + \ord{L^{-2}}\,.
\end{eqnarray}
Absorbing the divergent term in $\lambda_\infty$, we recover the vacuum flows which start from $\rho\to\infty$ at $\lambda\to-\infty$. In contrast to this, the Ricci flow  \eqref{eq:ricci-flow-L1} starts from a finite value $\lambda = \lambda_\infty$. Therefore, this is no longer the ancient solution problem. The endpoint, if the flow exists up to there, varies depending on $k$. For $k=0,-1$, it can flow to $\lambda \to \infty$, but for $k= 1$ it can only flow to $\lambda\to \lambda_\infty + L^2/2$.\footnote{This discussion applies to $L^2>0$. For a positive cosmological constant, the results for $k=1$ and $k=-1$ are exchanged.} It should be interesting to study further these possibilities. Let us note that AdS geometries of this kind have been investigated recently in \cite{Aharony:2019vgs}. Ref.~\cite{Emparan:2011ve} described exact horizon mergers in dS .

\subsection{Matter fields}
If there are matter fields one obtains coupled flow equations. For example, with a minimally coupled massless scalar field $\phi$, the Einstein equations give
\begin{eqnarray}
\fr{\rho} \partial_\rho g_{ab} = 2 R_{ab} - \partial_a \phi \partial_b \phi\,,
\end{eqnarray}
($\phi$ is appropriately scaled to appear in the leading order equation) and the Klein-Gordon equation reduces to a heat equation
\begin{eqnarray}
 \fr{\rho}\partial_\rho \phi + \nabla^2 \phi=0\,.
\end{eqnarray}


\section{Outlook}\label{sec:outlook}

The large-$D$ expansion has allowed us to give a very complete and explicit description of the topology-changing merger transition in the black hole/black string system, with simple solutions for the geometries at both sides of the transition ---the fused and split conifolds--- which extend far from the high-curvature region near the singularity and connect to the complete horizon of the localized black hole/non-uniform black string.

A remarkable aspect of the analysis is the relation to non-linear flow equations and solutions that have been studied in mathematics: the Ricci flow equation and the logarithmic diffusion equation. The possibility that this finding has deeper meanings or implications is tantalizing. 

Our study demonstrates the adaptability of the method of large-$D$ expansion. By considering different scaling regimes, it can cover all the aspects of the phases of the black hole/black string system: from weakly to strongly non-uniform black strings  \cite{Emparan:2018bmi}, to the singular merger transition, and to the localized black hole phases. The study of the latter configurations trivializes when the black hole becomes smaller than the length of the compact circle by a difference larger than $\ord{1/D}$: in this case the black hole does not feel the gravitational effect of its images, and it is an undistorted Schwarzschild black hole. 

Therefore, the large-$D$ expansion does indeed capture all the regimes of the black hole/black string system. However, in this article we have not computed the thermodynamic properties that are needed to complete its phase diagram. This requires further work, which involves  higher order corrections in $1/D$ and post-Newtonian corrections to the interaction, and is left for the future. 

Conifold transitions are widely present, possibly universally, in topology-changing mergers in the solution space of black hole systems ---\eg besides the cases considered in \cite{Emparan:2011ve}, they are also expected in the transitions between black droplets and black funnels in AdS \cite{Marolf:2013ioa}, and perhaps also between black holes hovering above an AdS black brane and black mushrooms \cite{Horowitz:2014gva}. We have found critical, split and fused conifolds that smooth out and extend the singular cones $S^2\times S^n$. It would be very interesting to find similar conifold geometries for $S^p\times S^n$ beyond the smooth cones in section \ref{sec:smoothcones}. However, this may require a better understanding of Ricci flows in more than two dimensions.

Finally, the success of the large-$D$ expansion in this problem is encouraging for its application in other cases. In particular, ref.~\cite{Kol:2005vy} argued that Choptuik's critical collapse of a scalar field \cite{Choptuik:1992jv} is related, via double Wick rotation, to the black hole/black string merger transition. It seems natural to try to apply our methods to this problem, and the large-$D$ scalings that we have introduced are likely relevant here. Interestingly, a recent large-$D$ study in \cite{Rozali:2018yrv} of a related system (critical behavior of gravitating scalars) invoked a scaling of temporal gradients that would map to the same scaling along the horizon as in our work. It will be of considerable interest to further develop this line of investigation.

\section*{Acknowledgements}
Work supported by ERC Advanced Grant GravBHs-692951 and MEC grant FPA2016-76005-C2-2-P. RS was also supported by JSPS KAKENHI Grant Number JP18K13541.

\appendix

\section{Horizon embedding}\label{app:embedding}

The condition for embedding \eqref{eq:horizonds2} into \eqref{eq:flatembed} is
\begin{eqnarray}
Y(\rho)=\rho,\quad X_{\pm}'(\rho) = \pm  2  \rho \sqrt{\coth(2(\rho^2-\mu))},\label{eq:embedding-con}
\end{eqnarray}
where $\pm$ denote each side of the horizon.
The embedded surface reproduces that of the single black hole at $\mu \to -\infty$.

At $\rho\to\infty$, the embedding behavior is independent of $\mu$,
\begin{eqnarray}
 X_\pm(\rho) \simeq  \pm \rho^2,
\end{eqnarray}
which is the same as that of the single black hole horizon.
In contrast, the behavior near the pinch varies depending on the value of $\mu$,
\begin{eqnarray}
X_\pm(\rho) \simeq \left\{
\begin{array}{cc}
\pm \sqrt{2}\sqrt{\rho^2-\mu} & (\mu \geq 0,\quad \rho \to \sqrt{\mu}),\\
\pm  \sqrt{\cosh(-2\mu)}\rho^2&(\mu<0,\quad \rho \to 0).
 \end{array}\right.
\end{eqnarray}

For the split conifold ($\mu<0$), we also need the embedding condition for the $S^n$-axis between the two horizons,
\begin{align}
 dX_{\rm axis} = \left.\sqrt{V}\right|_{\rho=0} d\xi =\sqrt{\frac{\sinh(2|\mu|)}{\cosh(2|\mu|)+\cosh(2\xi)}}d\xi.
\end{align}
Integrating from $\xi=-\infty$ to $\xi=\infty$, we obtain the separation between the horizons on the axis
\beq
\Delta X_{\rm axis} (\mu) =\left. \int_{-\infty}^\infty d\xi \sqrt{V}\right|_{\rho=0} = 2 \sqrt{\tanh|\mu|} K(\tanh^2|\mu| )
\,.\label{eq:delzint}
\eeq
Here $K(x)$ is the complete elliptic integral of the first kind.
Multiplying by the appropriate prefactor, this gives \eqref{eq:delzexact}.

Finally, imposing the mirror symmetry $X_-(\rho) = -X_+(\rho)$, we obtain the complete expression for $X_\pm(\rho)$ (figure \ref{fig:embedding})
\begin{align}
X_\pm (\rho) = \left\{\begin{array}{cc} \pm F(\rho^2-\mu)&(\mu\geq 0), \\ \pm F(\rho^2-\mu)\mp F(-\mu) \pm \fr{2}\Delta X_{\rm axis}(\mu) & (\mu <0), \end{array}\right.
\end{align}
where $F(x)$ is given by the integration of (\ref{eq:embedding-con}),
\begin{eqnarray}
 F(x)=-\fr{2} \arctan\left(\sqrt{\coth(2x)}\right)
 +\fr{4}\ln \left(\frac{\sqrt{\coth(2x)}+1}{\sqrt{\coth(2x)}-1}\right)
 +\frac{\pi}{4}.
\end{eqnarray}
$F(x)$ is adjusted to be $F(0)=0$. Its asymptotic behavior at $x\to\infty$ is
\begin{eqnarray}
F(x) = x+\frac{\pi+\ln 4}{8}+\ord{e^{-4x}}.
\end{eqnarray}

\section{Double cone perturbation}\label{app:coneperturb}

Consider the King-Rosenau solution \eqref{eq:KR-sol} with $\mu>0$, and expand it for small values of $\rho^2-\mu$ to one higher order than \eqref{eq:merger-waist}, to find 
\begin{eqnarray}
 V \simeq \frac{\rho^2-\mu}{\cosh^2\xi}\left(1+\frac{2}{3}(\rho^2-\mu)^2P_2(\tanh(\xi))+\Ord{(\rho^2-\mu)^3}\right)\,.
 \label{eq:merger-sol-expand-waist}
\end{eqnarray}
The leading term yields the neck of fused and critical solutions, \eqref{eq:largeDcone-V}, with the $SO(3)\times SO(n+1)$ symmetry of the $S^2$ and $S^n$. The first correction, where $P_2(x)$ is the second Legendre polynomial, is a dipolar deformation of the $S^2$.

Perturbative deformations of the $S^2$ for the double cone metric, $\mu=0$, were studied in \cite{Asnin:2006ip}, which found them of the form
\begin{eqnarray}
\rho^{s^s_+} P_\ell(\tanh(\xi))\,,
\end{eqnarray}
with
\begin{eqnarray}
 s^s_+ = \fr{2}\left(2-D + \sqrt{(D-2)(4\ell^2+12\ell+D+6)}\right)\,.
\end{eqnarray}
When $D\to\infty$ we have
\begin{eqnarray}
 s^s_+ =(\ell+2)(\ell-1)+\ord{D^{-1}},
\end{eqnarray}
and since $\ell=1$ is a pure gauge mode, we see that our large-$D$ result correctly gives the leading $\ell=2$ mode perturbation.

\section{Perturbations of the Schwarzschild black hole}\label{app:schwpert}

\subsection{Large $n$ limit of $S^{n+1}$-harmonics}
Assuming $SO(n+2)$ symmetry, the scalar harmonics on $S^{n+1}$ satisfy
\begin{eqnarray}
 \Pi_\ell''(\theta) + n \cot\theta\, \Pi_\ell'(\theta) + \ell(\ell+n) \Pi_\ell(\theta) = 0.
 \label{eq:harmonics-eq}
\end{eqnarray}
This is solved by the Gegenbauer polynomials
\begin{eqnarray}
 \Pi_\ell (\theta) = C_\ell^\frac{n}{2}(\cos\theta).
\end{eqnarray}
To zoom in to the small polar cap, we define a local coordinate
\begin{equation}
\vrho=\sqrt{n} \sin\theta.
\end{equation}
Then, assuming $\hat{\ell}\equiv\ell/n =\ord{1}$, the large-$n$ limit of \eqref{eq:harmonics-eq} becomes
\begin{eqnarray}
 \fr{\vrho} \partial_\vrho \Pi_\ell  + \hat{\ell}(\hat{\ell}+1))\Pi_\ell \simeq 0,
\end{eqnarray}
which results in
\begin{eqnarray}
 \Pi_\ell(\cos \theta) \simeq \fr{\sigma_{\ell}} e^{-\frac{\hat{\ell}(\hat{\ell}+1)}{2}\vrho^2}
=\fr{\sigma_{\ell}} e^{-\frac{n\hat{\ell}(\hat{\ell}+1)}{2}\sin^2\theta}\label{eq:s-harmonics-largeD},
\end{eqnarray}
where the coefficient $\sigma_{\ell}$ is determined by the value at the pole ($\theta=0$).
This formula is valid only for 
\begin{equation}
 \vrho^2 = n\sin^2\theta \ll n.
\end{equation}
Using Stirling's formula, we can estimate the value at large $n$,
\begin{eqnarray}
\sigma_{\ell}\equiv\fr{C^\frac{n}{2}_\ell (1)} = \frac{\ell! \Gamma(n)}{\Gamma(\ell+n)} \simeq \sqrt{2\pi \hat{\ell}(\hat{\ell}+1)n}
\left[ \frac{\hat{\ell}^{\hat{\ell}}}{(\hat{\ell}+1)^{\hat{\ell}+1}}\right]^n \left(1+\ord{n^{-1}}\right).
\end{eqnarray}

The Gegenbauer polynomials satisfy the orthogonality relation
\begin{eqnarray}
\int_0^\pi \sin^n \theta C_\ell^\frac{n}{2}(\cos\theta) C^\frac{n}{2}_m(\cos\theta)d\theta
= \frac{\pi  \Gamma(\ell+n)}{2^{n-1}(\ell+n/2)\ell!\Gamma(n/2)^2} \delta_{\ell,m}.\label{eq:Pi-orthonomal}
\end{eqnarray}
For $\ell = n \gg 1$, Stirling's formula gives
\begin{eqnarray}
\left. \frac{\pi  \Gamma(\ell+n)}{2^{n-1}(\ell+n/2)\ell!\Gamma(n/2)^2}\right|_{\ell=n} \simeq \frac{4^n}{3\sqrt{2}n}\,.
\end{eqnarray}

\subsection{Scalar-derived vector and tensor perturbations}\label{sec:sn-harmonics-VT}
We decompose the metric of $S^{n+1}$ as
\begin{align}
\gamma_{IJ} dz^I dz^J = d\theta^2 + \sin^2\theta \omega_{ij} d\sigma^i d\sigma^j,
\end{align}
where $\omega_{ij}$ is the metric of $S^n$.
Given the scalar harmonics $\Pi_\ell (\theta)$,
we can construct scalar-derived vector and tensor harmonics
\begin{align}
&{\sf V}_{I}(\theta) = \cD_I \Pi_\ell(\theta),\\
&{\sf T}_{IJ}(\theta) =\left[\cD_I \cD_J-\fr{n+1}\gamma_{IJ} \right]\Pi_\ell(\theta),
\end{align}
where $\cD_I$ is the connection compatible with $\gamma_{IJ}$.

In the following, we consider the only relevant mode to our analysis, namely $\ell =n$.
Using \eqref{eq:s-harmonics-largeD}, we can derive the match for the scalar-derived vector and tensor perturbations,
\begin{align}
&{\sf V}_{\theta}(\theta) =  \partial_\theta \Pi_n (\theta)\simeq -\frac{2 \sqrt{n}}{\sigma_n} \vrho e^{-\vrho^2}, \\
&{\sf T}_{\theta\theta}(\theta) =\left[ \partial_\theta^2 - \frac{n}{n+1}\cot\theta \partial_\theta\right]\Pi_n(\theta)
\simeq \frac{4n}{\sigma_n} \vrho^2 e^{-\vrho^2},\\
&{\sf T}^i{}_j(\theta) =\fr{n+1}\cot\theta\partial_\theta \Pi_n(\theta) \delta^i{}_j
\simeq - \frac{2}{\sigma_n} e^{-\vrho^2}\delta^i{}_j.
\end{align}

Note that in order to perform the match of geometries we must use that ${\sf T}_{ij}$ is smaller than ${\sf T}_{\theta\theta}$ by $1/n$.

Eliminating the scalar-derived vector and tensor by using some gauge transformations: $x\to x+\delta x$ and $\theta \to \theta + \delta \theta$, 
Eq.~\eqref{eq:KR-match-BH-pert} reproduces the perturbative result for $\ell =n$~\cite{Gorbonos:2004uc,Gorbonos:2005px}.
However, the matching ladder will be different, because we are assuming the touching limit $L \simeq 2r_0$, instead of the small black hole limit $L\gg r_0$.

Multipoles do not contribute directly to the thermodynamical quantities, except through the monopole in the second order perturbation made of them.
Therefore, the effect from the conifold cap on the thermodynamics can be estimated by using the orthogonal relation for $\Pi_\ell(\theta)$~\eqref{eq:Pi-orthonomal},
\begin{equation}
(\delta h_{\mu\nu})^2 \sim \sigma_n^2 \Pi_\ell(\theta)^2 \sim \sigma_n + (\rm higher \ multipoles ).
\end{equation}
To extract here the monopole component in the square of $\Pi_n(\theta)$, we use
\begin{eqnarray}
e^{-2n\sin^2\theta} \simeq \sigma_n^2  \Pi_n(\theta)^2 \simeq \frac{\sqrt{2\pi}\sigma_n}{3\sqrt{n}} + ({\rm higher\  multipoles}).
\end{eqnarray}

\section{Large-$D$ scaling of perturbations around the neck}\label{app:pertneck}

Following \cite{Kol:2002xz} we consider the ansatz (with $n$ finite)
\begin{eqnarray}
ds^2 = d\rho^2 + P(\rho) d\Omega_2^2+Q(\rho)d\Omega_{n}^2\,,
\end{eqnarray}
and expand the double cone metric around $\rho=0$, assuming $P(\rho)=\ord{\rho^2}$ and $Q(\rho) = \ord{1}$.\footnote{One can also expand assuming $P(\rho)=\ord{1}$ and $Q(\rho) = \ord{\rho^2}$, which corresponds to split cones. However, as we saw, the large-$D$ limit in this case is smooth without the need of finer resolution.
}
Then we obtain
\begin{eqnarray}
P(\rho) = \rho^2 \exp\left(2\sum_{i=1}p_i (\rho/\rho_0)^{2i}\right),\quad Q(\rho)=\rho_0^2\exp\left(2\sum_{i=1}q_i (\rho/\rho_0)^{2i}\right)\label{eq:neck-exp-sols}
\end{eqnarray}
where
\begin{align}
&p_1= -\frac{n(n+1)}{36} ,\quad p_2=\frac{n(n+1)^2 (31 n+72)}{32400},\nonum
&q_1 = \frac{n+1}{6},\quad q_2=-\frac{1}{540} (n+1)^2 (2 n+9). \label{eq:neck-exp-cf}
\end{align}
Near $\rho=0$, introducing $X=\rho\cos \chi$ and $Y=\rho\sin \chi$, this metric has a regular neck
\begin{eqnarray}
 ds^2 &\simeq&  d\rho^2+\rho^2 (d\chi^2- \cos^2 \chi dt^2)+\rho_0^2 d\Omega_{n}^2\\
& =& dY^2+dX^2-X^2 dt^2+\rho_0^2d\Omega_{n}^2.
\end{eqnarray}
The important point for us is that in order to obtain a finite large-$D$ limit this series solution requires that we  rescale the coordinate
\beq
\bar\rho=n\rho\,.
\eeq
The metric components then become
\begin{eqnarray}
 d\rho^2 = \frac{d\bar{\rho}^2}{n^2},\quad P(\rho) = \frac{ {\bar\rho}^2 }{n^2}\bar{P}(\bar\rho),\quad Q(\rho) = \rho_0^2 \left(1+\ord{n^{-1}}\right),
\end{eqnarray}
where $\bar{P}(\bar\rho)$ is a function of $\bar\rho$.
This is the same scaling as in section \ref{subsec:nearneck}.

\section{Matching the fused conifold to the near-neck region}\label{app:matchnearneck}
Consider the near-neck limit of the fused conifold~(\ref{eq:sol-neck-nlo}).
Up to $\ord{n^{-2}}$, the solution becomes
\begin{align}
&S =\mu x \left(1+\frac{2\ln x+x}{nx}+\frac{x^2-4+4(1+2x)\ln x-2x \ln^2 x}{n^2x^2}+\ord{n^{-3}}\right),\\
&N = 1 + \frac{1}{n}\frac{1+x}{x}+\frac{2(x+1)(x+3-2\ln x)}{n^2 x^2}+\ord{n^{-3}},
\end{align}
where $x=\rho^2/\mu-1$.
With $\rho$ as in \eqref{eq:nearneckrho}, but now also assuming $1\ll \sigma\ll n$, the near-neck limit of the fused conifold becomes
\begin{subequations}\label{eq:neck-near-neck-exp}
\begin{align}
& S \simeq \frac{2\sigma \mu}{n} \left(1+\frac{\ln \sigma-\sigma_0}{\sigma}+\frac{\ln \sigma - \sigma_0-1}{\sigma^2}+\ord{\sigma^{-3},n^{-1}}\right),\label{eq:R-near-neck-exp}\\
& N^2 d\rho^2 \simeq \frac{\mu d\sigma^2}{n^2} \left(1 + \fr{\sigma}+\frac{3/2+\sigma_0-\ln \sigma}{\sigma^2}+\ord{\sigma^{-3},n^{-1}}\right)\,, \label{eq:B-near-neck-exp}
\end{align}
\end{subequations}
where $\sigma_0 = \ln(n/2)$. For $\ln n \ll n$, we treat $\sigma_0$ as a $\ord{1}$ constant. Since the $\ord{n^{-k}}$ correction has
at most the same order of the pole, we only need the terms up to $\ord{n^{-k}}$ to obtain the expansion up to $\ord{\sigma^{-k}}$.

This metric can be connected to the near-neck metric~(\ref{eq:near-neck-ansatz}).
The two coordinates are related through $d\varsigma = N(\sigma)d\sigma$, which gives
\begin{eqnarray}
 \varsigma = \sigma - \varsigma_0+ \fr{2}\ln \sigma+\frac{4\ln \sigma-4 \sigma_0-1}{8\sigma}+\ord{\sigma^{-2}},
\end{eqnarray}
where $\varsigma_0$ is a constant.
Substituting this in (\ref{eq:neck-alpha-exp-as}) and (\ref{eq:neck-beta-exp-as}), the near-neck solution can be rewritten in $\sigma$ as
\begin{subequations}\label{eq:near-neck-sols-as-sigma}
\begin{eqnarray}
 \alpha^2 \simeq 2\sigma \left(1+\frac{\ln \sigma-\varsigma_0}{\sigma}+\frac{\ln \sigma-(2+\varsigma_0+\sigma_0)/2}{\sigma^2} \right) \label{eq:near-neck-sols-alpha-sigma}
\end{eqnarray}
and
\begin{eqnarray}
 e^\frac{\beta}{n+1} \simeq  1 + \fr{n}\left(\sigma+\beta_0-\varsigma_0+\frac{\varsigma_0-\sigma_0}{2\sigma}\right).\label{eq:near-neck-sols-beta-sigma}
\end{eqnarray}
\end{subequations}
Comparing (\ref{eq:near-neck-sols-alpha-sigma}) with (\ref{eq:R-near-neck-exp}), and also (\ref{eq:near-neck-sols-beta-sigma}) with $\rho$ as in \eqref{eq:nearneckrho}, we complete the matching by setting\footnote{In the match, we assume $1\ll \sigma \ll n$, but the condition $\sigma \gg 1$ is only used for obtaining the asymptotic expansion in both solutions. It might be possible to do the matching using only the weaker condition $\sigma\ll n$.}
\begin{eqnarray}
\varsigma_0=\beta_0=\sigma_0=\ln(n/2).
\end{eqnarray}

\section{Ricci flows with a cosmological constant}\label{app:cosmoflow}

Consider the ansatz~\eqref{eq:ansatz-0}, replacing now $d\Omega_n^2$ with the metric $d\Sigma_{(k)n}^2$ for the maximally symmetric space with curvature $k=0,\pm 1$.
The evolution equation and the scalar constraint are\footnote{$\Lambda$ in \eqref{eq:lambda}
is at most $\ord{n^2}$. If we allowed $y$ dependence in $H$, the term $n^3 (\nabla H)^2$ might be balanced in the scalar constraint by a value $\ord{n^3}$ of $\Lambda$. We will not pursue this possibility here.}
\begin{eqnarray}
 \fr{N}\partial_\rho K^{A}{}_{B} = \bar{R}^{A}{}_{B}-KK^{A}{}_{B} +\frac{n+p}{L^2}\delta^{A}{}_{B}- \frac{1}{N}\nabla^A \nabla_B N,\label{eq:adm-evolve-L}
\end{eqnarray}
and
\begin{eqnarray}
 K^2-K^{A}{}_{B} K^{B}{}_{A} =R+\frac{(n+p)(n+p-1)}{L^2},
\end{eqnarray}
where $L$ is the curvature radius of a negative cosmological constant. For a positive cosmological constant, replace $L\to iL$.

Again, we solve the scalar constraint to restrict the gauge
\begin{eqnarray}
 0=K^2-K^{A}{}_{B} K^{B}{}_{A}-\bar{R}-\frac{(n+p)(n+p-1)}{L^2}  = n^2 \left[\frac{H'^2}{N^2} - \frac{k}{H^2}- \fr{L^2}\right] +\ord{n}.
 \end{eqnarray}
This requires that $N$ be a function of only $\rho$. Choosing $H(\rho)=\rho$ we obtain\footnote{
An alternative gauge choice is $N=1+\ord{n^{-1}}$, which leads to
\begin{eqnarray}
H(\rho) = \left\{ \begin{array}{cl}
L \sinh(\rho/L) & (k=1), \\
 e^{\rho/L}& (k=0),\\
 L \cosh(\rho/L)&(k=-1).
\end{array}\right.
\end{eqnarray}
The result for $k=1$ seems consistent with \cite{Aharony:2019vgs}.}
\begin{eqnarray}
 N = \fr{\sqrt{k+\rho^2/L^2}} \left(1+\frac{N_1(\rho,y)}{n}\right).
\end{eqnarray} 
We again gauge-fix $C=0$.
The evolution equation for $K^a{}_b$ yields \eqref{eq:ricci-flow-L0} and the conformally-related flow \eqref{eq:ricci-flow-L1}.

The leading order metric becomes
\begin{eqnarray}
ds^2 = \left(1+\frac{2N_1}{n}\right) N_0^2 d\rho^2 + \fr{n N_0^2}\tilde{g}_{ab}dy^a dy^b+\rho^2 d\Sigma_{(k)n}^2\,,
\end{eqnarray}
where 
\beq
N_0=\fr{\sqrt{k+\rho^2/L^2}} 
\eeq
and
$N_1$ is determined as
\begin{equation}
N_1 = \fr{2}\rho^2N_0^2 R = \fr{2} \rho^2 N_0^4 \tilde{R}.
\end{equation}

For $k=1$, one can check that the small black hole limit $\rho \ll L$ reproduces the vacuum behavior.

\end{document}